%% file: AWA.tex
\documentclass[journal]{IEEEtran}
\usepackage[T1]{fontenc}    
\usepackage{hyperref}       
\usepackage{url}            
\usepackage{booktabs}       
\usepackage{amsfonts}       
\usepackage{nicefrac}       
\usepackage{microtype}      

\usepackage{lipsum}		
\usepackage{amsmath}
\usepackage[font=small,labelfont=bf]{caption}
\usepackage[english]{babel}
\usepackage{algorithm}
\usepackage{algpseudocode}
\usepackage{float}
\usepackage{multirow}
\usepackage[table,xcdraw]{xcolor}
\usepackage{pgfplots}
\usepackage{tikz}           
\usetikzlibrary{matrix}
\usepgfplotslibrary{groupplots}
\pgfplotsset{compat=newest}
\pgfplotsset{grid style={dashed,gray!50}}

\usepackage{pgfplots}
\usepgfplotslibrary{groupplots}
\usetikzlibrary{pgfplots.groupplots}
\usetikzlibrary{plotmarks}
\usetikzlibrary{calc}
\usepgfplotslibrary{external}
\usepackage{caption}
\usepackage{subcaption}
\usepackage{graphics}
\usepackage{bbm}
\usepackage{svg}
\usepackage{pdfpages}
\usepackage{hhline}
\usepackage{ragged2e}

\newcommand{\comment}[1]{}

\tikzset{%
	base/.style = {rectangle, rounded corners, draw=black,
		minimum width=2cm, minimum height=1cm,
		text centered,},
	conv/.style = {base, fill=blue!30},
	deconv/.style = {base, fill=green!30},
	maxpool/.style = {base, fill=orange!30},
	dense/.style = {base, fill=yellow!30},
}

\ifCLASSINFOpdf
\else
\fi
\begin{document}
%
\title{AWA: Adversarial Website Adaptation}
%
%
%

\author{Amir~Mahdi~Sadeghzadeh,
	Behrad~Tajali,
	and~Rasool~Jalili
\thanks{
The authors are with the Department of Computer Enginnering,
	Sharif University of Technology, Iran,
	Tehran, 11365-11155 (e-mail: amsadeghzadeh@ce.sharif.edu; behradtajali@ce.sharif.edu; jalili@sharif.edu).
}

}

%
%

\markboth{IEEE TRANSACTIONS ON INFORMATION FORENSICS AND SECURITY, VOL. *, NO. *, MONTH YEAR}%
{Shell \MakeLowercase{\textit{et al.}}: Bare Demo of IEEEtran.cls for IEEE Journals}
%



\maketitle

\begin{abstract}
One of the most important obligations of privacy-enhancing technologies is to bring confidentiality and privacy to users' browsing activities on the Internet. The website fingerprinting attack enables a local passive eavesdropper to predict the target user's browsing activities even she uses anonymous technologies, such as VPNs, IPsec, and Tor. Recently, the growth of deep learning empowers adversaries to conduct the website fingerprinting attack with higher accuracy. In this paper, we propose a new defense against website fingerprinting attack using adversarial deep learning approaches called Adversarial Website Adaptation (AWA). AWA creates a transformer set in each run so that each website has a unique transformer.
{\color{black}Each transformer generates adversarial traces to evade the adversary's classifier. AWA has two versions, including Universal AWA (UAWA) and Non-Universal AWA (NUAWA). Unlike NUAWA, there is no need to access the entire trace of a website in order to generate an adversarial trace in UAWA. }
We accommodate secret random elements in the training phase of transformers in order for AWA to generate various sets of transformers in each run. We run AWA several times and create multiple sets of transformers. If an adversary and a target user select different sets of transformers, the accuracy of adversary's classifier is almost 19.52\% and 31.94\% with almost 22.28\% and 26.28\% bandwidth overhead in UAWA and NUAWA, respectively. If a more powerful adversary generates adversarial traces through multiple sets of transformers and trains a classifier on them, the accuracy of adversary's classifier is almost 49.10\% and 25.93\% with almost 62.52\% and 64.33\% bandwidth overhead in UAWA and NUAW, respectively.

\end{abstract}


%
\IEEEpeerreviewmaketitle

\section{Introduction}
\label{sec:introduction}
\input{content/introduction}

\section{Background}
\label{sec:Background}
\input{content/background}

\section{Related Works}
\label{sec:Related Works}
\input{content/related}

\section{Adversarial Website Adaptation}
\label{sec:AWA}
\input{content/proposed_approach}

\section{Evaluation}
\label{sec:Evaluation}
\input{content/evaluation}

{\color{black}
	\section{AWA In Practice}
	\label{sec:awaprac}
	\input{content/AWAPrac}
}

\section{Conclusion}
\label{sec:Conclusion}
\input{content/conclusion}

\section*{Acknowledgments}

The authors would like to express their very great appreciation to Dr. Mahdieh Soleymani Baghshah for her valuable discussions. They also would like to offer their special thanks to Saeed Shiravi and all the anonymous reviewers for their valuable reviews and feedback.


%




\bibliographystyle{IEEEtran}
\bibliography{IEEEabrv.bib,ref.bib}
%



%

\appendices
\section{}
\label{sec:appendixa}
\input{content/Apendix_A}



\begin{IEEEbiography}[{\includegraphics[width=1in,height=1.25in,clip,keepaspectratio]{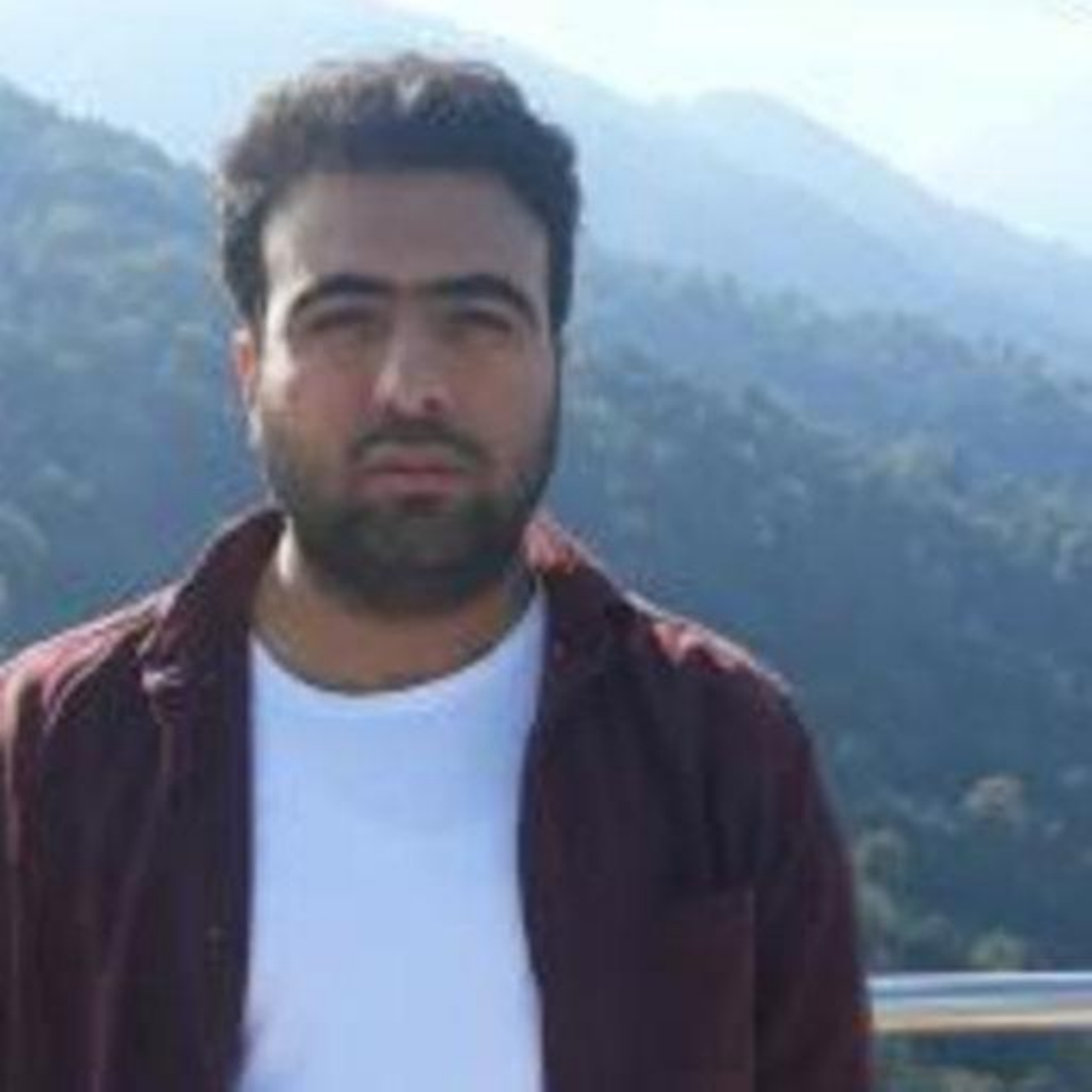}}]{Amir Mahdi Sadeghzadeh}
	received his B.Sc. degree in Information Technology Engineering from Isfahan University of Technology in 2014 and his M.Sc. degree in Computer Engineering from Sharif University of Technology in 2016. He is currently a Ph.D. candidate at the Department of Computer Engineering, Sharif University of Technology. His research interests include Deep Learning Security, Adversarial Deep Learning, and Privacy Enhancing Technologies.
	
\end{IEEEbiography}

\begin{IEEEbiography}[{\includegraphics[width=1in,height=1.25in,clip,keepaspectratio]{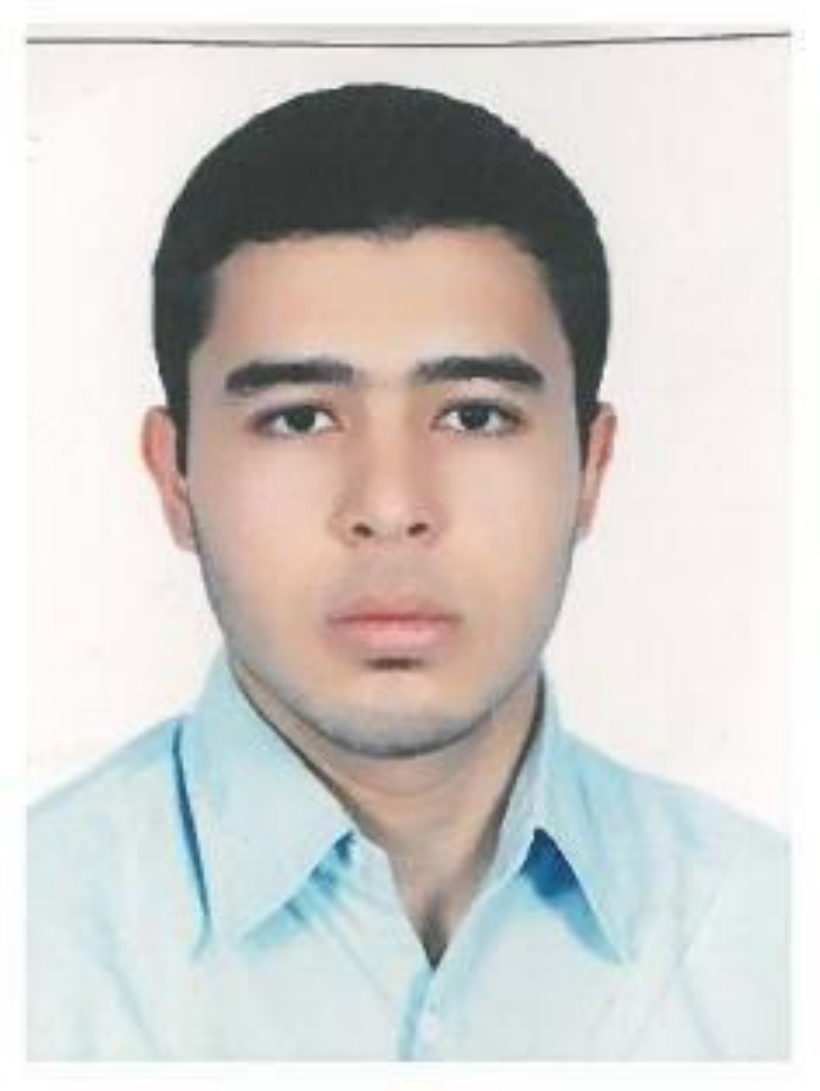}}]{Behrad Tajali}
	received his B.Sc degree from KN Toosi university of technology in 2016, and his M.Sc. degree from Sharif University of Technology in 2020 both majored in computer science. His personal research interests lie in Adversarial Machine Learning, Web Security \& Privacy, and Utilizing AI Algorithms in Security-sensitive Applications.
\end{IEEEbiography}

\begin{IEEEbiography}[{\includegraphics[width=1in,height=1.25in,clip,keepaspectratio]{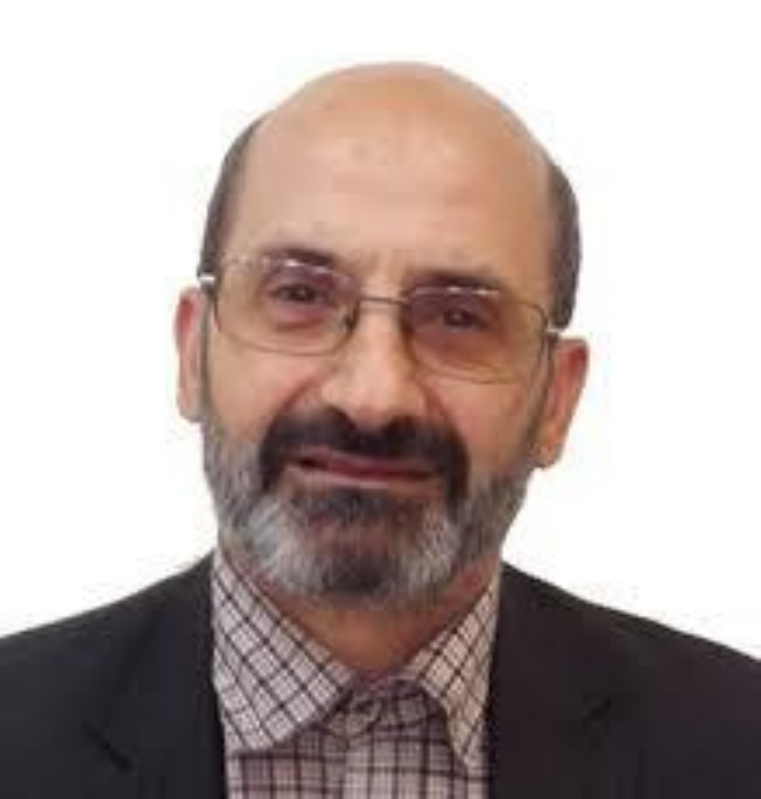}}]{Rasool Jalili}
	received his B.Sc. degree in Computer Science from Ferdowsi University of Mashhad in 1985, and his M.Sc. in Computer Engineering from Sharif University of Technology in 1989. He received his Ph.D. in Computer Science from The University of Sydney, Australia, in 1995. He then joined the Department of Computer Engineering, Sharif University of Technology, Tehran, Iran, in 1995. He has published more than 150 papers in Computer Security and Pervasive Computing in international journals and conferences proceedings. He is now an associate professor and the director of Data and Network Security Lab (DNSL) in Sharif University of Technology. His research interests include Access Control, Vulnerability Analysis, Database Security, and Machine Learning Security.
	
\end{IEEEbiography}


\end{document}

%% file: content/introduction.tex
\IEEEPARstart{W}{}ebsite fingerprinting (WF) attack is one of the most serious threats against the anonymity of the users' browsing activities. It enables an adversary to determine which website is visited by a target user even she uses Privacy Enhancing Technologies (PETs), such as VPNs, IPsec, and Tor \cite{DBLP:conf/ccs/HerrmannWF09, DBLP:conf/uss/HayesD16, DBLP:conf/ndss/PanchenkoLPEZHW16, DBLP:journals/popets/WangG16, DBLP:conf/ccs/SirinamIJW18, DBLP:conf/ndss/RimmerPJGJ18,DBLP:journals/tifs/ZhuoZZZZ18}. 
{\color{black}
The target user is a general term referring to a client-side victim whom the adversary intends to monitor and figure out which specific websites she has been visiting through her browser application.
	When an adversary visits a website through PETs, she can extract a unique trace associated with the statistical features of that website's network flow, such as the sequence of packets' direction. The adversary can visit various websites several times and collect a trace set from those. To conduct the website fingerprinting attack, the adversary can train a classifier on the collected trace set and uses this classifier to predict the website that has been visited by the target user. } The fundamental part of website fingerprinting attack is the classifier being used by the adversary to classify the target user's browsing activities. Recently, Deep Neural Networks (DNNs) have shown great performance in classifying network traces of various websites\cite{DBLP:conf/ccs/SirinamIJW18, DBLP:conf/ndss/RimmerPJGJ18}. Besides, DNNs-based classifiers have a highly better performance on defenses that has been proposed in the previous studies against website fingerprinting attack, such as WTF-PAD\cite{DBLP:conf/ccs/SirinamIJW18}. 

However, researchers have recently shown that DNNs have a serious vulnerability, which is called adversarial example \cite{DBLP:journals/corr/SzegedyZSBEGF13}. An adversarial example is a maliciously crafted input that causes classifiers to predict incorrectly. There are various methods to generate adversarial examples \cite{
	DBLP:journals/corr/SzegedyZSBEGF13,
	DBLP:journals/corr/GoodfellowSS14,
	DBLP:conf/sp/Carlini017,
	DBLP:conf/cvpr/Moosavi-Dezfooli17,
	ijcai2018-543}. 
{\color{black} Adversarial examples are considered as an attack to the classifiers in the literature of adversarial machine learning. However, it can be considered as a defense mechanism against the adversary's classifier in the website fingerprinting domain.
PETs can generate adversarial traces using conventional adversarial example generating methods, such as FGSM \cite{DBLP:journals/corr/GoodfellowSS14} and C\&W \cite{DBLP:conf/sp/Carlini017} to cause the adversary's classifier to predict incorrectly. However, there is a major challenge in using adversarial traces as a defense against website fingerprinting attack. 
Since PETs are publicly available in the threat model of website fingerprinting attack, an adversary can generate adversarial traces of various websites by her PETs and trains a classifier on them. Since training on adversarial examples, called adversarial training \cite{DBLP:journals/corr/GoodfellowSS14,DBLP:conf/iclr/MadryMSTV18}, is one of the most effective countermeasures against adversarial examples, the adversary's classifier being trained on adversarial traces can detect the true class of the target user's adversarial traces with a high success rate \cite{imani2019mockingbird,DBLP:conf/ndss/ZhangHRZ19}.} Therefore, we need an adversarial trace generating method, which is more resistant against adversarial training in the threat model of website fingerprinting attack.

We propose Adversarial Website Adaptation (AWA), which is a new defense against website fingerprinting attacks. AWA generates adversarial traces that are more resistant to adversarial training. 
{\color{black}The output of AWA is a transformer set so that each website has a unique transformer. Each transformer consists of a generator that generates adversarial perturbations, which are added to traces to make adversarial traces.
	We accommodate secret random elements in the training phase of AWA to create various sets of transformers in each run.
	When two transformer sets are created by various secret random elements, they generate adversarial traces with different distributions.
	The critical assumption of AWA is that the adversary knows that AWA generates the traces of the target user; however, she has no knowledge about the secret random elements of the target user's transformer set.}
We demonstrate when an adversary and a target user generate their adversarial traces by different transformer sets, the accuracy of adversary's classifier is low in classifying target user's adversarial traces. 
AWA has two versions, including Universal AWA (UAWA) and Non-Universal AWA (NUAWA). Transformers in UAWA use adversarial perturbations that are independent of website traces. Hence UAWA can generate adversarial traces on the fly, and there is no need to have access to the entire trace of a website before generating adversarial traces. Transformers in NUAWA use the entire trace of a website to generate adversarial traces.

 We run AWA several times and generate multiple sets of transformers to evaluate the performance of AWA. We assume an adversary randomly selects a set of transformers and generates adversarial traces of various websites through them, then she trains a classifier on them. The target user also randomly selects a set of transformers and generates adversarial traces of her browsing activities through them. The results demonstrate that if the transformer sets of an adversary and a target user are different, the accuracy of adversary's classifier is almost 19.52\% and 31.94\% with almost 22.28\% and 26.28\% bandwidth overhead in UAWA and NUAWA, respectively. We also evaluate the performance of AWA, encountering a powerful adversary that can generate adversarial traces of various websites through multiple sets of transformers and train a classifier on them. The results indicate that AWA must impose more bandwidth overhead to traces to decrease the accuracy of the adversary's classifier in this setting. UAWA and NUAWA decrease the accuracy of adversary's classifier to almost 49.10\% and 25.93\% with almost 62.52\% and 64.33\% bandwidth overhead, respectively.

The main contributions of this paper are as follows:
\begin{itemize}
	\item We propose AWA as a defense against website fingerprinting attack, which is more resistant against adversarial training. Also, AWA uses adversarial machine learning approaches and generates black-box adversarial traces.
	\item We present two versions of AWA, including universal AWA and non-universal AWA, and conduct multiple experiments to compare their performance.
	\item 
	We introduce the concept of secret random elements in the context of generating adversarial traces.
	\item 
We propose intra-class distance criterion, which is used to justify the effectiveness of AWA.
	
\end{itemize}

The rest of the paper is organized as follows. In Sec. \ref{sec:Background}, the preliminary and the threat model of website fingerprinting are described. Also, the preliminary of DNNs and basic methods for generating adversarial examples are introduced. Sec. \ref{sec:Related Works} reviews previous studies on website fingerprinting attacks and defenses. In Sec. \ref{sec:AWA}, AWA is presented. We also discuss the constraints of transformers in modifying network traces. 
Sec \ref{sec:Evaluation} evaluates the performance of UAWA and NUAWA in two scenarios.
Sec. \ref{sec:awaprac} discusses how to use AWA in practice. 
Lastly, in Sec. \ref{sec:Conclusion}, the conclusions of this study will be discussed.

%% file: content/background.tex
The Onion Router (Tor) is one of the most popular PETs, which provides anonymity and protection from eavesdropping. Different kinds of research have been conducted previously to breach the anonymity feature of Tor \cite{DBLP:conf/ccs/HerrmannWF09, DBLP:conf/uss/HayesD16, DBLP:conf/ndss/PanchenkoLPEZHW16, DBLP:journals/popets/WangG16, DBLP:conf/ccs/SirinamIJW18, DBLP:conf/ndss/RimmerPJGJ18,DBLP:journals/tifs/ZhuoZZZZ18}. Website fingerprinting is one such type of traffic analysis attack through extracting traffic patterns from the visited websites can form a unique fingerprint of acquired traffic flow and undermine Tor protection by detecting the website that the user has been just visiting.

\subsection{Website Fingerprinting (WF)}
Three entities play vital roles in the website fingerprinting attack, an adversary, a target user, and Privacy-Enhancing Technologies (PETs). A target user utilizes PETs to keep her communications confidential and anonymous. Since most PETs are publicly available, an adversary can use them to generate a set of network flows from her favorite websites and train a classifier on them to predict the target user's browsing activities. AS adversary has no access to the label of network flows being generated by the target user, she can not evaluate the performance of her classifier. A website fingerprinting defense mechanism aims to obfuscate the features from which the adversary's classifier learns to recognize and differentiate network flows of a website from those of others. 
{\color{black}
Likewise the previous studies, in this paper, we focus on Tor as the most popular PETs. However, AWA does not depend on the Tor architecture and can be generalized to any PETs.
}

		\begin{figure}[!t]
	\begin{center}
		\scalebox{1}{\includegraphics[width=1\linewidth]{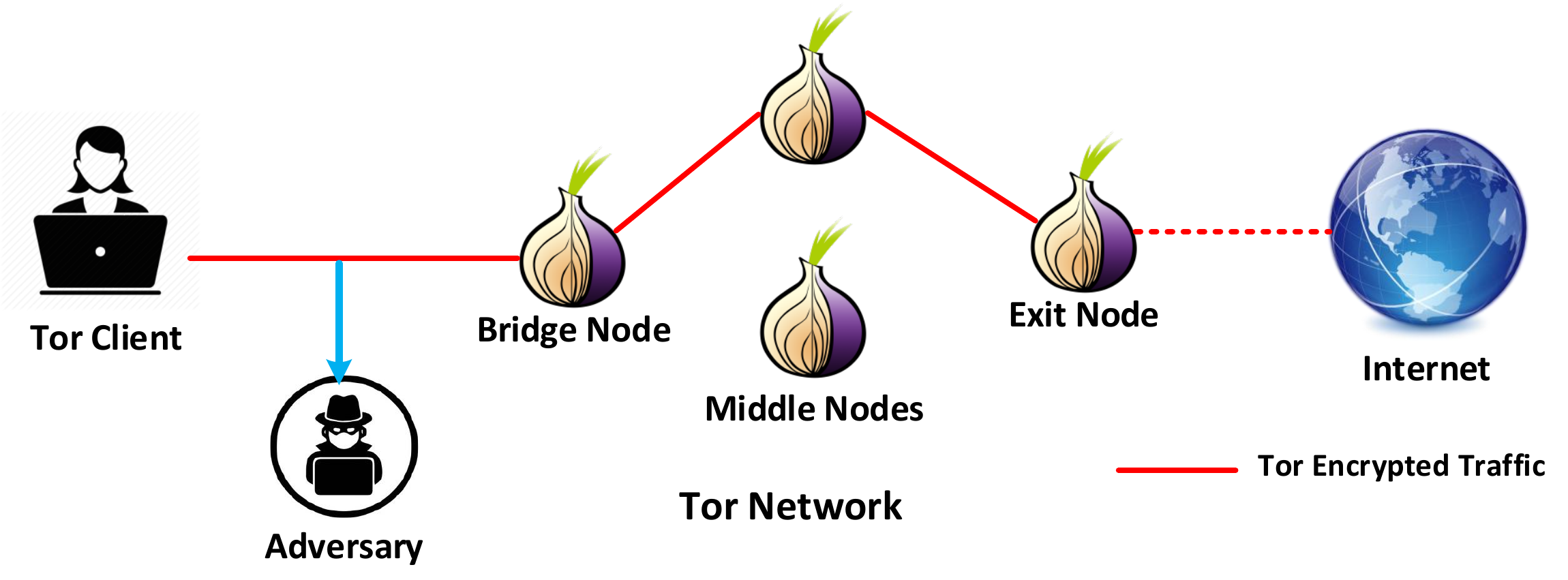}}
	\end{center}	
	\caption{A depiction for a typical WF attack scenario.}
	\label{FIG:Threat_model}
\end{figure}

\subsubsection{Threat Model}

Two initial presumptions are taken for the adversary in WF scenarios. First, the adversary is local, which means she is located somewhere between the Tor client and the Tor bridge  where she can access the link. Second, the adversary is passive, meaning that she could only capture the ongoing packets and is not allowed to manipulate, drop, or delay any of them. Any potential eavesdropper conforming such potential features can can play the role of the adversary (e.g., Autonomous Systems, Internet Service Providers, and network administrators). The adversary is also considered to be aware of the client's identity. A common assumption in the literature is that an adversary can parse the network flows of various websites, and isolate it from the other kinds of network traffic. 
{\color{black}
	Figure \ref{FIG:Threat_model} depicts the threat model of website fingerprinting attack.

	An essential point about any defense mechanism being
	implemented on PETs is that an adversary has access to
	network traffic being generated by the mechanism.
	For the success of an attack, the adversary needs to capture a set of network flows while visiting a collection of websites, notably those that she is interested to detect. Thereafter, some unique features shall be extracted from each network flow that helps each website become conspicuous to classifiers, amongst other websites. There are loads of such features in WF literature, such as packet size \cite{DBLP:conf/ccs/HerrmannWF09}, time and volume of traffic \cite{DBLP:conf/wpes/PanchenkoNZE11}, edit-distance score \cite{DBLP:conf/wpes/WangG13}, and the rate of traffic bursts in both directions \cite{DBLP:conf/uss/WangCNJG14}. By doing so, the adversary could attain several feature vectors that are used to train her supervised classifier. We consider each feature vector of a website's network flow as a website trace in the rest of this paper.
	
	There are two settings to evaluate WF attacks and defenses: closed-world and open-world.  The closed-world setting assumes that the target user only visits the websites belonging to a monitored set, and the adversary also uses the traces of the same set for training of her classifier.   In the more realistic open-world setting, the target user can visit arbitrary website regardless of the monitored set. Hence, the adversary needs to identify if a website belongs to the monitored set and to distinguish between monitored websites. Since the target user is restricted to visit the websites belong to the monitored set in the closed world setting, the adversary has an easier job of classifying her visited websites than in the open-world setting. 
	Hence, defeating the adversary in this setting would be considered as overcoming her in the most advantageous situation. We only evaluate AWA in the close-word setting, which is the worst setting for a defense mechanism.
}

\subsubsection{Website Trace Representation}

The communication of a client and a server is in the format of a bidirectional flow. A bidirectional flow is a sequence of packets that share the same source and destination IP addresses, port numbers, and protocol. Recent studies \cite{abe2016fingerprinting,DBLP:conf/ndss/RimmerPJGJ18,DBLP:conf/ccs/SirinamIJW18} have demonstrated that the direction of packets in bidirectional flows is enough for website fingerprinting attacks. Since recent attacks \cite{abe2016fingerprinting,DBLP:conf/ndss/RimmerPJGJ18,DBLP:conf/ccs/SirinamIJW18} and defenses \cite{imani2019mockingbird,nasr2020blind,DBLP:conf/infocom/0004SFCS19} have focused on the sequence of packets direction, we also use the same features in this study.
Each trace of a website is a sequence of packets direction in the order that they are received, which is called Direction Sequence (DS). For $DS^i$, we have:
\begin{equation}
\begin{split}
& DS^i =\; <d^i_1, d^i_2, ..., d^i_n> , \quad s.t. \quad d^i_j \in \{+1,-1\},\\
\end{split}
\end{equation}
where $n$ is the number of packets and $d^i_j$ is the $j^{th}$ packet direction of trace $i$. The direction of packets from the client to the server is $+1$ and from the server to the client is $-1$.
In this study, we consider each website trace as a Burst Sequence (BS) to simplify making perturbation for each trace. Burst is a sequence of consecutive packets all having the same direction. The burst sign is the sign of its including packets. BS is a sequence of bursts sizes multiplied to the bursts directions. For $BS^i$, we have:
{\color{black}
\begin{equation}
\begin{split}
& BS^i =\; <|B^i_1| \times D^i_1, |B^i_2| \times D^i_2, ..., |B^i_m| \times D^i_m>, \\ s.t. \;&B^i_j = < d^j_1, d^j_{2}, ..., d^j_{l}>,\:|B^i_j|=l,  \:D^i_j \in \{+1,-1\}, \:d^j_k = \\& D^i_j, D^i_{p} = -1 \times D^i_{p-1},\: k \in [1,l],\: j \in [1,m],\: p \in [2,m]
\end{split}
\end{equation}
where $m$ is the number of bursts, $B^i_j$ is the $j^{th}$ burst of $BS^i$, and the $D^i_j$ is the direction of this burst, which is the same as all packets in $B^i_j$. As an example, suppose that {\small$DS^i = <+1,-1,-1,-1,+1,+1,-1, -1>$}. The direction sequence $DS^i$ has four bursts, {\small$B^i_1 = <+1>$, $B^i_2 = <-1,-1,-1>$, $B^i_3 = <+1,+1>$,} and {\small$B^i_4 = <-1,-1>$}. Hence, {\small$|B^i_1|=1, D^i_1=+1$, $|B^i_2|=3, D^i_2=-1$, $|B^i_3|=2, D^i_3=+1$,} and {\small$|B^i_4|=2, D^i_4=-1$}. The burst sequence $i$ is {\small$BS^i = <1\times(+1),3\times(-1),2\times(+1),2\times(-1)> \:=\: <+1,-3,+2,-2>$}. In the rest of study, each website trace is a burst sequence, and a trace set is a set of burst sequences.
}

\subsection{Deep Neural Networks}
\label{lab:sgd}
Deep neural networks (DNNs) are multi-layer functions that receive $x_i \in \mathcal{X}$ as input and output $y_i = f_{DNN}(x_i)$ ($y_i \in \mathcal{Y}$). DNNs are revealed to have high performance on raw data extracted from network traffic and do not need manual feature engineerings to reach high accuracy. Three important DNNs which are mostly used in network traffic classification are Convolutional Neural Networks (CNN), Recurrent Neural Networks (RNN), and Stacked Denoising Autoencoders (SDAE).  Amongst which, One-Dimensional Convolutional Neural Network (1D-CNN) has got the best results in network traffic classification tasks \cite{DBLP:conf/ndss/RimmerPJGJ18,DBLP:journals/tnsm/AcetoCMP19, DBLP:conf/ccs/SirinamIJW18}. 
A neural network consists of three parts: an input layer, a bunch of hidden layers inside, and an output layer. 
{\color{black}Each layer includes a set of neurons with non-linear activation functions and parameters $\theta$ (weights and biases) related to them. Different architectures of deep neural networks have different numbers of parameters.
During the training phase, the optimum values for the network parameters are calculated so that the loss function $J$, which indicates the distance between actual and predicted labels in supervised classification, will be minimized:
$
\theta^{*} = \underset{\theta}{\mathrm{argmin}} \; J(\theta, x, y).
$
The optimization is usually done using some common versions of the Stochastic Gradient Descent (SGD). SGD is an iterative method, where $\theta$ is initialized randomly and is  updated iteratively using the gradient of the loss function with respect to the parameters in every iteration. The gradient is calculated using a mini-batch of training data. The size of a mini-batch is often 64, 128, or 256, and the data included in the batch is often randomly selected from training data. It is notable that putting different data in the mini-batch changes the gradients of the loss function with respect to parameters.
Therefore, $\theta^*$  depends on the initialization of  $\theta$ and the order of the training data which SDG is confronted with in every mini-batch. }

\subsection{Adversarial Examples}
In spite of their remarkable achievements and high performance in complicated tasks, DNNs have been demonstrated having a critical vulnerability \cite{DBLP:journals/corr/SzegedyZSBEGF13}. {\color{black}Adversarial examples are intentionally crafted inputs, which cause the victim classifier $f$ to make a mistake about the correct label of the input.} Considering $f^*$ as a classifier that always designates the correct label, we define adversarial example $x'$ as:
\begin{equation} 
\begin{split}
&f^*(x') = y, \quad f(x') = y',\quad s.t.\; y'\, \neq y\\
\end{split}
\end{equation}
The typical approach to make adversarial example $x'$ is adding the adversarial perturbation vector $\eta$ to the real input $x$ like: $x' = x + \eta$. After crafting $x'$, the actual class of $x'$ and $x$ should be the same. 
\comment{
In some domains such as image classification, the amount of perturbation should be limited to a specific bound, such as $P$-norm $\left\lVert . \right\rVert_p$ since it is obvious that a large amount of perturbation or noise could easily change the true class of data. The $P$-norm is defined as follows:
\small
\begin{equation}
\left\lVert x'-x \right\rVert_p =\biggl (\sum_{i=0}^{D-1}|x'_i-x_i|^p\biggr )^{1/p}
\end{equation}
\normalsize
where $D$ is the dimension of data $d$ and $p \in [0,\infty]$. Accordingly, the adversarial example generating process could be formulated as follow:
\small
\begin{equation}
\begin{split}
\label{AE_MP}
& \underset{x'}{\mathrm{argmin}} \;\left\lVert x' - x \right\rVert_p\\
s.t.\: & f^*(x') = y, f(x') = y', y' \neq y, x' \in Domain(x)
\end{split}
\end{equation}
\normalsize
}
Several methods have been introduced to generate adversarial examples so far, of which Fast Gradient Sign Method (FGSM) \cite{DBLP:journals/corr/GoodfellowSS14} and Carlini and Wagner (CW) \cite{DBLP:conf/sp/Carlini017} are two popular ones. 
Moosavi-Dezfooli \textit{et al.} \cite{DBLP:conf/cvpr/Moosavi-Dezfooli17} proposed universal adversarial perturbations for the first time where the adversary creates adversarial perturbation independent of a particular input. In this attack, a universal adversarial perturbation is added to a set of samples and cause the target classifier to predict the label of most of samples wrongly. 

{ \color{black}
Adversarial examples are considered as an attack to classifiers in the literature of adversarial machine learning. However, it can be considered as a defense mechanism against the adversary's classifier in the website fingerprinting domain. Tor can implement adversarial example generating methods such as FGSM \cite{DBLP:journals/corr/GoodfellowSS14} and C\&W \cite{DBLP:conf/sp/Carlini017} and use them to perturb traces of various websites and generate adversarial traces to cause the adversary's classifier to predict incorrectly. Such a defense mechanism does not work because of the threat model of website fingerprinting attacks.
Since Tor is publicly available, an adversary can generate adversarial traces of various websites and train a classifier on them. In the adversarial machine learning literature, training on adversarial examples is one of the most effective defenses against adversarial example attacks \cite{DBLP:journals/corr/GoodfellowSS14,DBLP:conf/iclr/MadryMSTV18}. 
Therefore, when an adversary trains a classifier on the adversarial traces, she is doing adversarial training, and her classifier becomes more robust against adversarial traces. Accordingly, where Tor uses adversarial example generating methods as a defense mechanism against adversary's classifier, adversarial training is a part of the threat model of website fingerprinting attack. Saidur \textit{et al.} \cite{imani2019mockingbird} and Zhang \textit{et al.} \cite{DBLP:conf/ndss/ZhangHRZ19} demonstrate that, because of adversarial training, adversarial traces being generated by FGSM \cite{DBLP:journals/corr/GoodfellowSS14} or C\&W \cite{DBLP:conf/sp/Carlini017} are not effective against the adversary's classifier. We propose AWA as a new defense mechanism against website fingerprinting attacks that generates adversarial traces which are more resistant against adversarial training. 
}

\subsection{Maximum Mean Discrepancy (MMD)}
Given two sets of data $X=\{x_1, ...,x_n\}$ and $Y=\{y_1, ..., y_m\}$, drawn identical independent distribution (i.i.d.) from distributions $P$ and $Q$ respectively, MMD criterion empirically estimates the distance between $P$ and $Q$ in Reproducing Kernel Hilbert Space (RKHS). An RKHS $H_k$ is a space of functions with a kernel $k: \mathcal{X}\times\mathcal{X} \rightarrow \mathbb{R}$ such that $f(x)=<f,k(.,x)> \forall f \in H_k$. In other words, MMD considers the distance between the embedded mean of two distributions as the distance between them.
MMD is defined as follows:
\begin{equation}
\begin{split}
MMD(X,Y) = \left\| 
\frac{1}{n} \sum_{i=1}^{n} \phi(x_i) - \frac{1}{m} \sum_{i=1}^{m}\phi(y_i)
\right\|_{\mathcal{H}_k}
\end{split}
\end{equation}
where $\mathcal{H}$ is a universal RKHS, $\phi(.) \in \mathcal{H}$ is the mapping of input space $\mathcal{X}$ to the RKHS, and $k(.,.) = <\phi(.),\phi(.)>$ is the universal kernel associated with this mapping. MMD can be easily approximated by sampling from distributions P and Q. We use MMD with a Gaussian kernel to estimate the distance between two distributions of websites.

%% file: content/related.tex
This section reviews the most prominent website fingerprinting attacks and defenses presented so far. 

\subsection{Website Fingerprinting Attacks}
Before the outbreak of deep learning, most of the studies used to apply manual feature engineering and conventional machine learning classifiers.
The first WF attack is presented by Hermann \textit{et al.} \cite{DBLP:conf/ccs/HerrmannWF09}, where they reach just 3\% accuracy for the closed world scnario containing 775 websites. Following that, Panchenko \textit{et al.} \cite{DBLP:conf/wpes/PanchenkoNZE11} achieve 55\% accuracy with enhanced extracted features on the same dataset. WF attacks gradually improve up to 90\% success rate using edit-distance \cite{DBLP:conf/ccs/CaiZJJ12, DBLP:conf/wpes/WangG13}. 
Wang \textit{et al.} \cite{DBLP:conf/uss/WangCNJG14} using a k-nearest neighbor classifier on a combination of features achieve 91\% accuracy in the closed-world setting containing 100 websites.
Panchenko \textit{et al.} \cite{DBLP:conf/ndss/PanchenkoLPEZHW16} introduce an attack based on a Support Vector Machine (SVM). For the closed-world setting, their method achieves 91\% accuracy.
Hayes and Danezis \cite{DBLP:conf/uss/HayesD16} suggest k-Fingerprinting attack (k-FP)
The k-FP attack achieves 91\% accuracy for closed-world setting.
Zhuo \textit{et al.} \cite{DBLP:journals/tifs/ZhuoZZZZ18} propose a new attack using Profile Hidden Markov Models. This attack 
achieves 99\% accuracy on SSH and Shadowsocks network traffic in the closed world setting.

Abe and Goto \cite{abe2016fingerprinting} introduce the first WF attack employing DNNs. They use a Stacked Denoising Autoencoder (SDAE) classifier, and instead of whole handcrafted features, used in previous studies, they feed their model just with a sequence of raw packets direction. They report 88\% accuracy on the closed-world scenario. 
Rimer \textit{et al.} \cite{DBLP:conf/ndss/RimmerPJGJ18} propose Automated Website Fingerprinting (AWF) attack, which utilizes multiple DNN structures, including SDAE, CNN, and LSTM to classify the sequence of packets direction for various website. According to the obtained results, their CNN model outperforms others with 96\% accuracy for the closed-word scenario.
Sirinam \textit{et al.} \cite{DBLP:conf/ccs/SirinamIJW18} develop a new CNN classifier which surpasses all previous ones and achieves more than 98\% accuracy in closed-word settings. The authors collect a dataset for 100 websites to evaluate their classifier. They claim that the sequence of packets direction is enough for classifying websites traces with high accuracy. Furthermore, their attack shows high performance against traces being protected by WTF-PAD \cite{DBLP:conf/esorics/JuarezIPDW16} and Walkie-Talkie \cite{DBLP:conf/uss/WangG17} methods.
Bhat \textit{et al.} \cite{DBLP:journals/popets/BhatLKD19} using the ResNet-18 architecture \cite{7780459} introduce VAR-CNN that could gain 98.8\% accuracy in closed-world settings. VAR-CNN uses the sequence of packets direction, inter-arrival time, and cumulative statistical features to classify traces. 

\subsection{Website Fingerprinting Defenses}
One of the early defenses against website fingerprinting attack is proposed by Dyer \textit{et al.} in \cite{DBLP:conf/sp/DyerCRS12} called Buffered Fixed-Length Obfuscator (BuFLO). 
BuFLO imposes high bandwidth and latency overhead on traces. Later, some extended versions of BuFLO, such as Tamaraw \cite{DBLP:conf/ccs/CaiNWJG14} and CS-BuFLO \cite{DBLP:conf/wpes/CaiNJ14} are proposed to solve this problem. 
By employing the Adaptive Padding method \cite{DBLP:conf/esorics/ShmatikovW06}, Juarez \textit{et al.} introduce WTF-PAD \cite{DBLP:conf/esorics/JuarezIPDW16}. WTF-PAD tries to obfuscate the inter-arrival time feature by filling the abnormal time gap between two packets of a sequence with dummy ones. 
Nithyanand \textit{et al.} in \cite{DBLP:conf/wpes/NithyanandCJ14} and Wang \textit{et al.} in \cite{DBLP:conf/uss/WangCNJG14} almost concurrently propose the idea of finding a representation for all traces that are close to each other and this representation is called super trace \cite{DBLP:conf/wpes/NithyanandCJ14} or  supersequence \cite{DBLP:conf/uss/WangCNJG14}. Both defenses use clustering to find traces that are close to each other. Then, every trace must be padded so that it becomes similar to the supersequence of the cluster to which it belongs. 
Wang \textit{et al.} in another study \cite{DBLP:conf/uss/WangCNJG14} present Walkie-Talkie (WT) that has two features: Burst-molding and half-duplex communications. 
WT has 31\% and 34\% bandwidth and latency overhead, respectively. 
Since Walkie-Talkie works on the half-duplex links, it has some issues in practice.


Some new methods recently have employed adversarial machine learning approaches to develop a defense against website fingerprinting attack. By leveraging the adversarial example notion, Saidur \textit{et al.} \cite{imani2019mockingbird} develop their defense method called Mockingbird. The defense procedure starts with selecting a target trace from a pool and calculating the gradient of the distance between the source trace and the target trace. Then it tries to move toward the target iteratively until the detector is deceived. Mocking Bird can reduce the accuracy of Deep Fingerprinting classifier \cite{DBLP:conf/ccs/SirinamIJW18} down to 35.2\% with almost 56\% bandwidth overhead.
{\color{black}
Zhang \textit{et al.} \cite{DBLP:conf/ndss/ZhangHRZ19} propose two defenses against video fingerprinting attacks using differential privacy called FPA and $d^*$. 
FPA and $d^*$ impose 200\% and 600\% bandwidth overhead. FPA decreases the accuracy of the target classifier from 94\% to almost 20\%. 
As the authors  mentioned in their paper, unlike streaming traffic, HTTP traffic is more interactive. Therefore, the proposed defenses are not applicable in the website fingerprinting domain.
}
Nasr \textit{et al.} \cite{nasr2020blind} propose Blind Adversarial Network Perturbations (BANP). The method injects universal perturbations into the traffic stream, leveraging remapping functions. The authors indicate that Deep Fingerprinting \cite{DBLP:conf/ccs/SirinamIJW18} attack has 8\% accuracy against BANP, and their method imposes 11.11\% bandwidth overhead on traces.
Abusnaina \textit{et al.} \cite{DBLP:conf/infocom/0004SFCS19} present Deep Fingerprinting Defender (DFD) approach. DFD  consists of two modules: the burst observer and the injection module. It works by injecting dummy packets dependent on the passing burst in either one-way or two-way operation mode. With 14.26\% bandwidth overhead, they decrease the accuracy of Deep Fingerprinting \cite{DBLP:conf/ccs/SirinamIJW18} attack to about 7.29\%. Although BANP \cite{nasr2020blind} and DFD \cite{DBLP:conf/infocom/0004SFCS19} evaluate their methods when the adversary is aware of the defense, the procedures of generating adversarial traces for adversary and target user are different in their evaluation. Since we can not choose which versions of PETs belong to adversary and target user in the threat model of website fingerprinting attack, the procedures of generating adversarial traces for them  must be the same.

%% file: content/proposed_approach.tex
We introduce Adversarial Website Adaptation (AWA) as a new defense mechanism against website fingerprinting attacks that generate adversarial traces which are more resistant against adversarial training. 
We assume statistical features of each website W has a unique distribution $D_W$, and traces of website W come from this distribution. 
Also, we have access to the trace set $TS$ and label set $Y_{TS}$, including traces and labels of all monitored websites. Trace set of website W is called $ TS_W $, and it can be used to estimate the distribution of website W (empirical distribution $ \hat{D}_W $). 
Each website has a unique transformer in AWA, and the goal of a transformer is to transform the distribution of the associated website to adapt to the transformed distribution of another website. 
Each transformer has a generator that generates adversarial perturbation to change a website trace, and since the size of adversarial perturbation controls the magnitude of bandwidth overhead, it must be minimized.

\input{content/old_AWA_overview.tex}

AWA has two versions, including Non-Universal AWA (NUAWA), and Universal AWA (UAWA). NUAWA needs to have access to the entire trace of a website before generating adversarial perturbations. 
UAWA uses universal adversarial perturbations and does not need to access the entire trace of a website before generating adversarial perturbations. Recently, Sadeghzadeh \textit{et al.} \cite{sadeghzadeh2020adversarial} and Nasr \textit{et al.} \cite{nasr2020blind} have demonstrated that using universal adversarial perturbation is effective in evading DNNs-based network traffic classifiers. In practice, UAWA generates a set of universal perturbations for each website, and whenever a user wants to visit a website, the pre-made perturbation of that website is added to the trace of user on the fly.
The only difference between NUAWA and UAWA is in their generators' inputs. They are fed by the trace set $ TS $ in NUAWA and the noise set $ Z $ in UAWA. The whole process of transforming the traces of website $ W $ is denoted as $T_W(TS_W, Z)$ where $ T_W $ is the transformer of website $ W $, $ TS_W $ is the trace set of website $W$, and $Z$ is a noise set.

{ \color{black} We explain AWA for a pair of websites and then extend it to K websites. AWA has a framework that changes the distributions of a pair of websites. 
 Suppose that website A has been paired with website B, and the distributions of websites A and B are $D_A$ and $D_B$, respectively. The transformers $T_A$ and $T_B$ change the distributions $D_A$ and $D_B$ to $D'_A$ to $D'_B$, respectively, so that $D'_A$ and $D'_B$ become close together. However, the size of change in distributions should be limited due to bandwidth overhead. }
Therefore, the AWA framework goal is to minimize the distance between $T_A(TS_A, Z)$ and $T_B(TS_B, Z)$ with the minimum amount of change on traces to minimize bandwidth overhead.  We utilize parameterized generators $G_A$ and $G_B$ in $T_A$ and $T_B$, respectively, and a parameterized discriminator $D_{AB}$ to minimize the distance between $T_A(TS_A,Z)$ and $T_B(TS_B,Z)$. The discriminator $D_{AB}$ wants to differentiate between $T_A(TS_A, Z)$ and $T_B(TS_B, Z)$. On the contrary, $ T_A $ and $ T_B $ want to prevent $D_{AB}$ from differentiating between them. 
{\color{black}
When the discriminator fails to differentiate between $T_A(TS_A,Z)$ and $T_B(TS_B,Z)$, the distributions $D'_A$ and $D'_B$ are adapted.}
We use an Auxiliary Classifier $AC$ to make transformers move traces of both classes, not just one of them. Auxiliary Classifier is a simple classifier that has been trained on the traces of all websites before transformation.
In order to apply the AWA framework for K classes of websites, the following three phases are introduced
\begin{itemize}
	\item 
	\textbf{Pre-training phase:} Auxiliary classifier is trained on the network traces of $K$ websites.
	\item 
	\textbf{Training phase:} First, $K/2$ pairs of websites are randomly selected. Then, generators and discriminator of the proposed framework are trained for each pair on the pre-collected set of network traces. The output of this phase is a transformer set consists of $K$ transformers for each website.
	\item 
	\textbf{Testing Phase:} New clean traces of each website are transformed by the respective transformer.
\end{itemize}
AWA can be run several times to create multiple sets of transformers.
Figure \ref{FIG:AWA_framework} shows the overview of AWA framework in training and testing phases for pair websites A and B.

\subsection{Transformers}

There are three constraints to transform traces of a website.
\begin{enumerate}
	\item
	
	Transformers can not remove any packets from traces; otherwise, the functionality of network traffic is disrupted. Hence, adding dummy packets is the only way to transform traces.
	\item
	As mentioned in section 2, we consider traces of each website as a burst sequence. The burst sequence consists of integer numbers, and it is discrete. The output of transformers must be integer numbers.
	\item
	If a burst in a trace is broken into two bursts by inserting a dummy packet having the opposite direction in the middle, the latency overhead is imposed on traces. Each breaking of a burst imposes two round trip times overhead on a trace. Since it is not intended to impose latency overhead on traces, transformers only add dummy packets at the end of bursts in the same direction.
\end{enumerate}

Each transformer consists of a generator, which feeds using a noise set or a trace set and generates a perturbation vector. The elements of perturbation vector specifies how many dummy packets are added to each burst of input trace. The first element of perturbation vector is added to the first positive burst of trace, and if a trace starts with a negative burst, the perturbation vector is shifted one element to the right. The value of perturbation vector is all positive, and it is multiplied by the sign of the target trace. Then it is added to the target trace to preserve the constraints 1 and 3.  As mentioned, the input of generators is different in NUAWA and UAWA. In NUAWA, the input of a generator is a trace set , and in UAWA, a generator is fed by a noise set. Although there is no need to the noise set $Z$ in NUAWA, we use the same notation  $T_W(TS_W,Z)$ for transformers in NUAWA and UAWA for simplicity. For $T_W$ we have:
\begin{equation}
\begin{split}
\text{NUAWA:}&\; T_W(TS_W,Z) = G(TS_W) \times sign(TS_W) + TS_W \\
\text{UAWA:}&\; T_W(TS_W,Z) = G(Z) \times sign(TS_W) + TS_W
\end{split}
\end{equation}
Although the output of transformers is not rounded in the training phase, it is rounded in the testing phase to preserve constraint 2.
{\color{black}
	The amount of BandWidth Overhead (BWO) that $T_W$ adds to a trace $tr$ is defined as follows:
	\begin{equation}
	BWO(\%) =  \frac{\parallel |T_W(tr,Z)| - |tr|\parallel_1}{\parallel |tr| \parallel_1}  * 100
	\end{equation}
	where $\parallel |x| \parallel_1$ is the sum of absolute values of $x$.
}

\subsection{Loss Functions}
The AWA framework has three kinds of loss functions to optimize the parameters of generators, discriminator, and auxiliary classifier. 
{\color{black}
	The discriminator's goal is to predict the label of generators correctly. However, the generators' goal is to bring close the distributions of their outputs together to confuse the discriminator and evade the auxiliary classifier. It is improper that generators add a lot of dummy packets to traces to change their distributions. Hence, bandwidth overhead loss restricts generators not to adding high bandwidth overhead to traces.}
The loss function of the auxiliary classifier is cross-entropy, which is a standard supervised learning loss, and is as follows:
\begin{equation}
\underset{\theta_{AC}}{\mathrm{min}} \: \mathcal{L}_{AC}(TS, Y_{TS}) \!= \!-\mathbb{E}_{(x,y) \sim(TS,Y_{TS})}\! \sum_{k=1}^{K} \! \mathbb{I}_{(y=k)} \! \log AC(x)_k
\end{equation}
where $K$ is the number of websites in $TS$,  $\mathbb{I}_{(y=k)}$ is one if $y=k$; otherwise is zero, $AC(.)_k$ is the $k^{th}$ element of AC output, and $\theta_{AC}$ is the parameters of AC. The output of discriminator is the label of the transformer from which the input of discriminator has been generated. Because there are only two transformers, their labels are zero and one. The loss function of $D_{AB}$ is defined as follows:
\begin{equation}
\begin{split}
\underset{\theta_{D_{AB}}}{\mathrm{min}} \; &\mathcal{L}_{D_{AB}}(TS_{A},TS_{B},\mathcal{P}) =\\ &-\mathbb{E}_{x_A \sim TS_A, z \sim \mathcal{P}} [\log D_{AB}(T_A(x_A,z))] \\
&- \mathbb{E}_{x_B \sim TS_B, z \sim \mathcal{P}} [\log ( 1 - D_{AB}(T_B(x_B,z)))]
\end{split}
\end{equation}
where $\theta_{D_{AB}}$ is the parameters of discriminator $D_{AB}$ and $\mathcal{P}$ is a random noise distribution. The loss function of generators consists of three parts. For simplicity, we only introduce the loss function of $G_A$, and the loss function of $G_B$ can be defined in the same way. $\mathcal{L}_{G_{A}\_AC}$ is the value of AC logits for the true class of the output of transformer. 
Logits is the output of the Penultimate layer in DNNs.
The purpose of this loss is to make the AC predict the label of the output of $T_A$ wrongly. We have:
\begin{equation}
\begin{split}
\mathcal{L}_{G_{A}\_AC}(TS_A, &Y_{TS_A},\mathcal{P}) = \mathbb{E}_{(x,y) \sim(TS_A,Y_{TS_A}), z \sim \mathcal{P}} \\
&\sum_{k=1}^{K} \mathbb{I}_{(y=k)} max(logits_{AC(T_A(x,z))_k}, 0)
\end{split}
\end{equation}
where $logits_{AC(T_A(x,z))_k}$ is the $k^{th}$ element of the output of $AC$ logits, and $Y_{TS_A}$ is the label of website A.   $\mathcal{L}_{G_{A}\_OH}$ regulates the magnitude of change that transformers add to traces. The magnitude of change between the input and output of transformers is the bandwidth overhead that AWA imposes on traces, which must be minimized. We have:
{\color{black}
	\begin{equation}
	\begin{split}
	&\mathcal{L}_{G_{A}\_OH}(TS_A, \mathcal{P},\tau_{low},\tau_{high}) =
	\mathbb{E}_{x \sim TS_A, z \sim \mathcal{P}}\; max(  \\
	&\frac{\parallel |T_A(x,z)| - |x| \parallel_1}{\parallel |x|\parallel_1} -\tau_{high},0) -min(\frac{\parallel |T_A(x,z)| - |x|\parallel_1}{\parallel |x|\parallel_1}\\ &-\tau_{low},0)
	\end{split}
	\end{equation}
	where $\tau_{high}$ and $\tau_{low}$ determine the level of bandwidth overhead inducing no penalty, and $\parallel |.| \parallel_1$ is the sum of absolute values.} $\tau_{high}$ controls the upper bound of bandwidth overhead, and if overhead is more than $\tau_{high}\times 100$, $\mathcal{L}_{G_{A}\_OH}$ increases. Since sometimes some generators do not move websites distributions, we add a loss term that makes generators move the distributions of websites. This term using  $\tau_{low}$ makes generators to add some bandwidth overhead to traces, and it increases if overhead is less than $\tau_{low} \times 100$.
{ \color{black}
We use the domain confusion objective that has been proposed by Tzeng \textit{et al.} \cite{DBLP:conf/iccv/TzengHDS15} as the third part of $G_A$ loss function. The discriminator's output is a value between zero and one, and when the output is closer to zero, the discriminator's prediction is the transformer with label zero; otherwise, the discriminator's prediction is the transformer with label one. 
When the distributions of transformers' output are very close together, the discriminator fails to differentiate between them, and its output becomes 0.5. 
Therefore, this part of the loss function makes transformers change their outputs so that discriminator's output become 0.5. The domain confusion loss function is defined as follows:}
\begin{equation}
\begin{split}
&\mathcal{L}_{G_{A}\_DC}(TS_A,\mathcal{P}) = - \mathbb{E}_{x \sim TS_A, z \sim \mathcal{P}}\\
& [\frac{1}{2}\log D_{AB}(T_A(x,z)) + \frac{1}{2}\log(1-D_{AB}(T_A(x,z)))]
\end{split}
\end{equation}
The complete loss of $G_A$ is as follows:
\begin{equation}
\begin{split}
&\underset{\theta_{G_{A}}}{\mathrm{min}} \; \mathcal{L}_{G_{A}}(TS_A,\mathcal{P},\tau_{low},\tau_{high},\alpha,\beta,\gamma) = \\
&\alpha\; \mathcal{L}_{G_{A}\_AC} + \beta\;  \mathcal{L}_{G_{A}\_OH} + \gamma\; \mathcal{L}_{G_{A}\_DC}
\end{split}
\end{equation}
where $\theta_{G_{A}}$ is the paprameters of $G_A$, and $\alpha$, $\beta$, and $\gamma$ regulate the impact of each part of loss in $\mathcal{L}_{G_{A}}$.

{\color{black}
	
\subsection{Secret Random Elements}
Although we move the distributions of $K/2$ pairs of websites towards each other, because of the limitation on bandwidth overhead being imposed on traces, websites distributions do not become indistinguishable. AWA must add a lot of bandwidth overhead to the traces of paired websites in order to become indistinguishable. This is impractical in the real world. Suppose that we run AWA once and create a transformer set S to generate adversarial traces, and also website A has been paired by website B in the training phase of AWA. $T^S_A$ and $T^S_B$ are the transformers of websites A and B in S, respectively. $T^S_A$ changes the distribution of website A from $D_A$ to $D'_A$, and $T^S_B$ also changes the distribution of website B from $D_B$ to $D'_B$. Since websites A and B were paired in the training process of S, the goal of AWA's framework is to bring $D'_A$ and $D'_B$ close together. If an adversary has no access to AWA, she only can train a classifier on samples of $D_A$ and $D_B$.
In this setting, if the target user has access to AWA and generates her adversarial traces with $D'_A$ and $D'_B$ distributions, the adversary's classifier being trained on the samples of $D_A$ and $D_B$ has low accuracy on the target user traces due to the auxiliary classifier in the AWA's framework.

If we put S in publicly available PETs, an adversary also can generate adversarial traces of different websites through S and train a classifier on them. The target user also uses PETs and generates her adversarial traces through S. In this setting, the adversary has access to $D'_A$ and $D'_B$, and she can train a classifier on samples of them. Since $D'_A$ and $D'_B$ are not indistinguishable and the target user also generates her traces based on these distributions, the adversary's classifier being trained on samples of $D'_A$ and $D'_B$ has high accuracy on the target user's adversarial traces. 

When an adversary trains a classifier on adversarial traces, she is doing adversarial training.
Zhang \textit{et al.} \cite{DBLP:conf/iclr/ZhangCSBDH19} demonstrate that adversarial training performance strongly correlates with the distance between the distribution of test data and the distribution of training data. Test samples that are far from the distribution of training data are more likely to be classified wrongly. 
In the website fingerprinting threat model, the training data of the adversary's classifier and also its test data (which is in fact the traces obtained from target user) are generated by the defense mechanisms of PETs. AWA as a defense mechanism is able to control both the training and the test data distributions of the adversary's classifier.
Therefore,  if AWA generates adversarial traces with different distributions for the adversary and the target user, the adversary's adversarially trained classifier would fail and get low accuracy on the target user's adversarial traces. However, there are multiple users in the real world, and we do not know which one is an adversary and which one is a benign user. Therefore, AWA must generate adversarial traces of each user by a unique distribution so that when a user, which can be an adversary, trains a classifier on her adversarial traces, can not classify other users' adversarial traces with a promising success rate.
With this purpose, AWA must create multiple sets of transformers so that each transformer set generates adversarial traces with different distribution from others. Also, an individual user should not be aware of what transformer set other users have chosen. Otherwise, an adversary can generate her adversarial traces by the same transformer set that the target user has picked up, and in this setting, the accuracy of the adversary's classifier is high.

We accommodate secret random elements in AWA to achieve these aims by inspiring the concept of keys in the cryptography literature. Based on the Kerckhoffs' principle \cite{kerckhoffs1883cryptographic}, it is assumed the encryption scheme is known to the adversary, but the key itself remains secret. 
Here in the website fingerprinting threat model, we have a similar assumption that PETs are publicly available, and thus the adversary is able to generate adversarial traces of any chosen website.
However, there is no secret key in the defenses against website fingerprinting attack. 
Notably, the functionality of keys in the cryptography field and secret random elements are very different. The only similarities between these two concepts are randomness and secrecy.
Secret random elements make AWA generate a different set of transformers in each run.
Suppose that we run AWA twice with various random elements to create two transformer sets $S_1$ and $S_2$.  An adversary uses $S_1$, and a target user uses $S_2$. As an example, the adversary changes the distribution of website A using $T^{s_1}_A$ from $D_A$ to $D^{s_1}_A$, and the target user changes the distribution of website A using $T^{s_2}_A$ from $D_A$ to $D^{s_2}_A$. If $D^{s_1}_A$ and $D^{s_2}_A$ are far enough from each other, the adversary's classifier being trained on the samples of $D^{s_1}_A$ reaches a low accuracy on the target user's adversarial traces being generated by $D^{s_2}_A$. 
It is noteworthy to mention that because of the bandwidth limitation $D^{s_1}_A$ and $D^{s_2}_A$ can not be very far from $D_A$.

As mentioned in section \ref{lab:sgd}, there are multiple random elements in the Stochastic Gradient Descent (SGD) algorithm, such as parameter initialization and the order of training set. SGD is used to minimize loss functions of generators and discriminators in the AWA training phase. We use the initial parameters of generators and discriminators, as well as the order of training set as the two parts of random elements. It is notable that since we have to use SGD in the training phase of transformers anyway, we do not add any new randomness to transformers. We just consider the initial parameters of generators and discriminators, as well as the order of training set as the random elements and keep them secret.

‌Besides, with respect to the capabilities of AWA's framework, we use the pair list and noise being fed to generators in UAWA as the third and fourth parts of the secret random elements. The pair list is a list of pairs with size $K/2$, determining which website is paired by which website in the AWA training phase. 
For example, suppose that we have four websites, A, B, C, and D. The pair list can be \{(A, C),(B, D)\}, which means that A is paired with C and B is paired with D in the AWA training phase.
When the  pair lists of an adversary and a target user are different, the distribution of the adversarial traces that the adversary collects and trains a classifier on differs from the adversarial traces that the target user generates. 
For example,  website A is paired with website B in the training phase of the adversary's transformer set, and website A is paired with website C in the training phase of the target user's transformer set. 
Hence, the adversary's classifier is more likely to be vulnerable to the target user's adversarial traces generated by different pair lists.
}
\input{content/AlgAWA.tex}

\subsection{AWA Algorithm}
The complete training process of AWA is presented in Algorithm \ref{alg:AWA}. 
After selecting a list of pairs, two transformers are trained in $T$ iterations for each pair of websites. In each iteration, each generator is trained in $ G_T $ iterations, and the discriminator is trained two times between the generators training in $ D_T$ iterations. Sometimes during the training of transformers in some iterations, they impose high bandwidth overhead to traces, which is unacceptable. Therefore we check the magnitude of bandwidth overhead during the training phase, and we only choose transformers that have reasonable bandwidth overhead. We specify a parameter $ OH $ and only select transformers with bandwidth overhead less than $ OH $ during the training. However, if the bandwidth overhead of transformers in all iterations during the training is more than $OH$, we select the transformers in the last iteration.
The output of the algorithm is a transformer set, which consists of $K$ transformers.

%% file: content/old_AWA_overview.tex
\begin{figure*}[t!]
	\centering
	\begin{subfigure}[t]{0.6\textwidth}
		\centering
		\resizebox{1\textwidth}{!}{%
			\begin{tikzpicture}[roundnode/.style={circle, draw=green!60, fill=green!5, very thick, minimum size=7mm},]
			\filldraw[color=black!20, fill=black!5] (0.4,0.1) rectangle  (6.1,2)  ;
			\filldraw[color=black!20, fill=black!5](0.4,-1.1) rectangle  (6.1,-3);

			\draw[->, very thick] (4,2.5)node[] [above] (TextNode1) {$TS_A$} -- (4,1.25) ;
			\draw[thick,dashed]  (4,2.25) --(-0.5,2.25) ;
			\draw[thick,dashed]  (-0.5,2.25) --(-0.5,1.3);
			\draw[->,thick,dashed]  (-0.5,1.3) -- (1,1.3) node[midway] [above] {\footnotesize if NUAWA};

			\draw[->, very thick] (4,2.5-3.2)node[] [above] (TextNode1) {$TS_B$} -- (4,1.25-3.2) ;
			\draw[thick,dashed]  (4,2.25-3.2) --(-0.5,2.25-3.2) ;
			\draw[thick,dashed]  (-0.5,2.25-3.2) --(-0.5,1.3-3.2);
			\draw[->,thick,dashed]  (-0.5,1.3-3.2) -- (1,1.3-3.2) node[midway] [above] {\footnotesize if NUAWA};
			
			\filldraw[color=black!5, fill=black!5, very thick] (5.,0.2) rectangle  (5.4,0.5) node[midway,black] {$T_A$};
			\filldraw[color=black!5, fill=black!5, very thick] (1.8,0.85) rectangle  (2.3,1.2) node[midway,black] {$G_A$};

			\draw[very thick] (1,0.3) -- (1,1.8);
			
			\draw[very thick] (1,0.3) -- (2,0.55);
			\draw[very thick] (2,0.55) -- (3,0.3);
			\draw[very thick] (1,1.8) -- (2,1.55);
			\draw[very thick] (2,1.55) -- (3,1.8);
			
			\draw[very thick] (3,0.3) -- (3,1.8);

			\draw (4,1.05) circle (2mm) node {$*$};
			\draw[->,thick, dashed] (-0.5,0.8) node[] [left] (TextNode1) {$Z$}-- (1,0.8) node[midway] [below] {\footnotesize if UAWA};
			\draw[->,very thick] (3,1.05) -- (3.8,1.05);
			\draw[->,very thick] (4.2,1.05) -- (6.1,1.05)node[midway] [above] (TextNode1) {$T_A(TS_A,Z)$};
			

			\filldraw[color=black!5, fill=black!5, very thick] (5.,0.2-3.1) rectangle  (5.4,0.5-3.1) node[midway,black] {$T_B$};
			\filldraw[color=black!5, fill=black!5, very thick] (1.8,0.85-3.2) rectangle  (2.3,1.2-3.2) node[midway,black] {$G_B$};

			\draw[very thick] (1,0.3-3.2) -- (1,1.8-3.2);
			
			\draw[very thick] (1,0.3-3.2) -- (2,0.55-3.2);
			\draw[very thick] (2,0.55-3.2) -- (3,0.3-3.2);
			\draw[very thick] (1,1.8-3.2) -- (2,1.55-3.2);
			\draw[very thick] (2,1.55-3.2) -- (3,1.8-3.2);
			
			\draw[very thick] (3,0.3-3.2) -- (3,1.8-3.2);

			\draw (4,1.05-3.2) circle (2mm) node {$*$};
			\draw[->,thick, dashed] (-0.5,0.8-3.2) node[] [left] (TextNode1) {$Z$}-- (1,0.8-3.2) node[midway] [below] {\footnotesize if UAWA};
			\draw[->,very thick] (3,1.05-3.2) -- (3.8,1.05-3.2);
			\draw[->,very thick] (4.2,1.05-3.2) -- (6.1,1.05-3.2)node[midway] [above] (TextNode1) {$T_B(TS_B,Z)$};
			
			
			\filldraw[color=black!60, fill=black!5, very thick](9,0.5) rectangle  (10,-3.5) node[midway] [rotate=90] (TextNode1) {$D_{AB}$};
			
			\draw[->, very thick] (6.1,1.05) -- (9,-1.5) node[] [left] (TextNode1) {};
			\draw[->, very thick] (6.1,1.05-3.2) -- (9,-1.5) node[] [left] (TextNode1) {};

			\filldraw[color=black!60, fill=black!5, very thick](9,0.7) rectangle  (10,2.5) node[midway] [rotate=90] (TextNode1) {$AC$};
			
			\draw[->, very thick] (6.1,1.05) -- (9,1.6) node[] [] (TextNode1) {};
			\draw[->, very thick] (6.1,1.05-3.2) -- (9,1.6) node[] [] (TextNode1) {};

			\draw[->, very thick] (10,-1.5) -- (11,-1.5) node[] [right] (TextNode1) {Transformer Label};
			\draw[->, very thick] (10,1.6) -- (11,1.6) node[] [right] (TextNode1) {Class Label};
			
			\end{tikzpicture}
		}%
		\caption{Training Phase}
	\end{subfigure}%
	~ 
	\begin{subfigure}[t]{0.4\textwidth}
		\centering
				\resizebox{1.0\textwidth}{!}{%
			\begin{tikzpicture}[roundnode/.style={circle, draw=green!60, fill=green!5, very thick, minimum size=7mm},]
			\filldraw[color=black!20, fill=black!5] (0.4,0.1) rectangle  (4.8,2)  ;
			\filldraw[color=black!20, fill=black!5](0.4,-1.1) rectangle  (4.8,-3);
			
			\draw[->, very thick] (4,2.5)node[] [above] (TextNode1) {$new\_trace_A$} -- (4,1.25) ;
			\draw[thick,dashed]  (4,2.25) --(-0.5,2.25) ;
			\draw[thick,dashed]  (-0.5,2.25) --(-0.5,1.3);
			\draw[->,thick,dashed]  (-0.5,1.3) -- (1,1.3) node[midway] [above] {\footnotesize if NUAWA};

			\draw[->, very thick] (4,2.5-3.2)node[] [above] (TextNode1) {$new\_trace_B$} -- (4,1.25-3.2) ;
			\draw[thick,dashed]  (4,2.25-3.2) --(-0.5,2.25-3.2) ;
			\draw[thick,dashed]  (-0.5,2.25-3.2) --(-0.5,1.3-3.2);
			\draw[->,thick,dashed]  (-0.5,1.3-3.2) -- (1,1.3-3.2) node[midway] [above] {\footnotesize if NUAWA};
			
			\filldraw[color=black!5, fill=black!5, very thick] (4.2,0.2) rectangle  (4.6,0.5) node[midway,black] {$T_A$};
			\filldraw[color=black!5, fill=black!5, very thick] (1.8,0.85) rectangle  (2.3,1.2) node[midway,black] {$G_A$};

			\draw[very thick] (1,0.3) -- (1,1.8);
			
			\draw[very thick] (1,0.3) -- (2,0.55);
			\draw[very thick] (2,0.55) -- (3,0.3);
			\draw[very thick] (1,1.8) -- (2,1.55);
			\draw[very thick] (2,1.55) -- (3,1.8);
			
			\draw[very thick] (3,0.3) -- (3,1.8);

			\draw (4,1.05) circle (2mm) node {$*$};
			\draw[->,thick, dashed] (-0.5,0.8) node[] [left] (TextNode1) {$Z$}-- (1,0.8) node[midway] [below] {\footnotesize if UAWA};
			\draw[->,very thick] (3,1.05) -- (3.8,1.05);
			\draw[->,very thick] (4.2,1.05) -- (5.5,1.05)node[] [right] (TextNode1) {$T_A(new\_trace_A,Z)$};
			

			\filldraw[color=black!5, fill=black!5, very thick] (4.2,0.2-3.1) rectangle  (4.6,0.5-3.1) node[midway,black] {$T_B$};
			\filldraw[color=black!5, fill=black!5, very thick] (1.8,0.85-3.2) rectangle  (2.3,1.2-3.2) node[midway,black] {$G_B$};

			\draw[very thick] (1,0.3-3.2) -- (1,1.8-3.2);
			
			\draw[very thick] (1,0.3-3.2) -- (2,0.55-3.2);
			\draw[very thick] (2,0.55-3.2) -- (3,0.3-3.2);
			\draw[very thick] (1,1.8-3.2) -- (2,1.55-3.2);
			\draw[very thick] (2,1.55-3.2) -- (3,1.8-3.2);
			
			\draw[very thick] (3,0.3-3.2) -- (3,1.8-3.2);

			\draw (4,1.05-3.2) circle (2mm) node {$*$};
			\draw[->,thick, dashed] (-0.5,0.8-3.2) node[] [left] (TextNode1) {$Z$}-- (1,0.8-3.2) node[midway] [below] {\footnotesize if UAWA};
			\draw[->,very thick] (3,1.05-3.2) -- (3.8,1.05-3.2);
			\draw[->,very thick] (4.2,1.05-3.2) -- (5.5,1.05-3.2)node[] [right] (TextNode1) {$T_B(new\_trace_B,Z)$};
			

			\end{tikzpicture}
		}%
		\caption{Testing Phase}
	\end{subfigure}
	\caption{\footnotesize An overview of Adversarial Website Adaptation (AWA) framework for two websites $A$ and $B$. 
		In the AWA training phase, the generators $ G_A $ and $ G_B $ are trained to make the transformers' outputs indistinguishable for discriminator $ D_{AB} $ and decrease the logits value of auxiliary classifier $ AC $ for the true class of traces. 
		The inputs of transformers are noise set $ Z $ in UAWA and trace sets $ TS_A $ and $ TS_B $ in NUAWA. In the testing phase, the new traces $ new\_trace_A $ and $ new\_trace_B $ are transformed by trained transformers $ T_A $  and $ T_B $, respectively.}
	\label{FIG:AWA_framework}
\end{figure*}
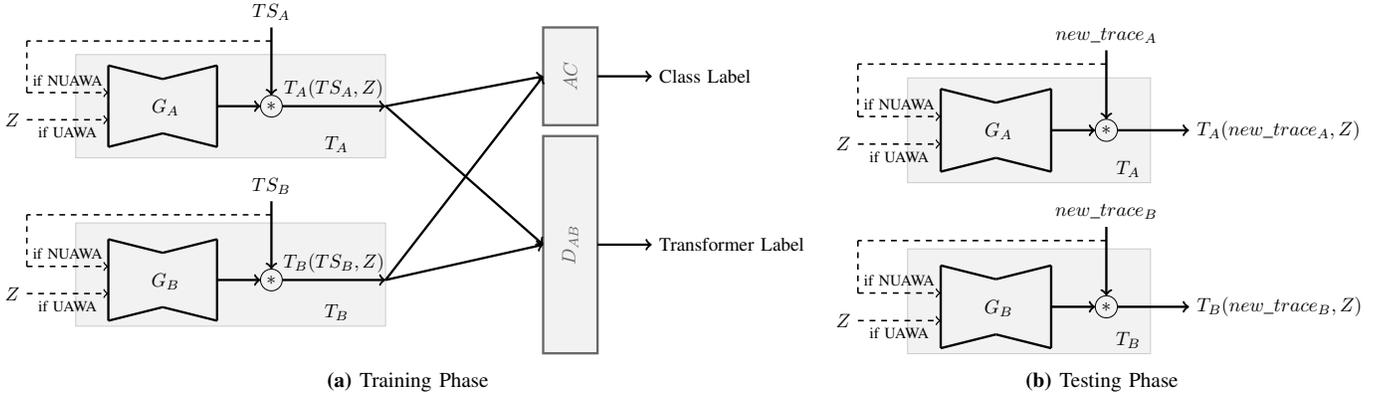

%% file: content/AlgAWA.tex
\begin{algorithm}[!t]
	\scriptsize
	\caption{Adversarial Website Adaptation (AWA)}
	\label{alg:AWA}
	\hspace*{\algorithmicindent} \textbf{Input:} Trace set $TS$, label set $Y_{TS}$, number of websites $K$, noise distribution \\ 
	\hspace*{\algorithmicindent}  $\mathcal{P}$, batch size $bs$, number of discriminator iterations $D_T$, number of\\
	\hspace*{\algorithmicindent} 
	 generator iterations $G_T$, overhead threshold $OH$, overhead controllers $\tau_{low}$\\
	 \hspace*{\algorithmicindent} 
	  and $\tau_{high}$,  parameters $\alpha, \beta, \gamma$, and number of training iterations $T$. \\
	\hspace*{\algorithmicindent} \textbf{Output:} Transformer set $S$
	\begin{algorithmic}[1]
		\State $AC \gets \text{train auxiliary classifier on}\; (TS,Y_{TS}) \;\text{using} \;\mathcal{L}_{AC}(TS,Y_{TS})$
		\State $Pairs \gets \text{randomly select} \;K/2 \;\text{pairs among websites}$
		\State $S \gets \text{empty transformer set with size}\; K$
		\For{$pair \;\textbf{in} \; Pairs$}
		\State $TS_A \gets \text{traces of website}\; pair[0]\; \text{in}\; TS$
		\State $TS_B \gets \text{traces of website}\; pair[1]\; \text{in}\; TS$
		\State $\text{randomly initialize parameters of}\; T_A,\; T_B,\; D_{AB}$
		\State	$selected\_T_A \gets \emptyset$
		\State	$selected\_T_B \gets \emptyset$
		\For{$t\gets 0, T$}	
				\For{$i\gets 0, D_T$}
				\State $TB_A \gets \text{randomly select}\; bs\;\text{traces from}\; TS_A$
				\State $TB_B \gets \text{randomly select}\; bs\;\text{traces from}\; TS_B$
				\State $\text{update} \;\theta_{D_{AB}}\; \text{to minimize}\; \mathcal{L}_{D_{AB}}(TB_{A},TB_{B}, \mathcal{P})$
		\EndFor
		\For{$i\gets 0, G_T$}
				\State $TB_A \gets \text{randomly select}\;bs\; \text{traces from}\; TS_A$
				\State $\text{update} \;\theta_{G_{A}}\; \text{to minimize}\; \mathcal{L}_{G_{A}}(TB_A, \mathcal{P},\tau_{low},\tau_{high},\alpha,\beta,\gamma)$			
		\EndFor
		\For{$i\gets 0, D_T$}
				\State $TB_A \gets \text{randomly select}\; bs\;\text{traces from}\; TS_A$
				\State $TB_B \gets \text{randomly select}\; bs\;\text{traces from}\; TS_B$
				\State $\text{update} \;\theta_{D_{AB}}\; \text{to minimize}\; \mathcal{L}_{D_{AB}}(TB_{A},TB_{B}, \mathcal{P})$
		\EndFor
		\For{$i\gets 0, G_T$}
				\State $TB_B \gets \text{randomly select}\; bs\; \text{traces from}\; TS_B$
				\State $\text{update} \;\theta_{G_{B}}\; \text{to minimize}\; \mathcal{L}_{G_{B}}(TB_B, \mathcal{P},\tau_{low},\tau_{high},\alpha,\beta,\gamma)$			
		\EndFor
		\If {$\frac{\parallel |T_A(TS_A,Z \sim \mathcal{P})| - |TS_A|\parallel_1}{\parallel |TS_A| \parallel_1}\hspace{-0.1cm}\leq \hspace{-0.05cm}OH \textbf{and}$ \\
			\quad \quad \quad \quad \quad \quad \quad \quad \quad \quad \quad \quad \quad \quad \quad $\frac{\parallel |T_B(TS_B,Z\sim \mathcal{P})| - |TS_B| \parallel_1}{\parallel |TS_B| \parallel_1} \hspace{-0.1cm}\leq \hspace{-0.05cm}OH$}
				\State	$selected\_T_A \gets T_A$
				\State	$selected\_T_B \gets T_B$
		\ElsIf{ $t = T-1\,\textbf{and}\, selected\_T_A = \emptyset\; \textbf{and}\;selected\_T_B = \emptyset $}	
				\State	$selected\_T_A \gets T_A$
				\State	$selected\_T_B \gets T_B$
		\EndIf
		\EndFor
		\State $S[pair[0]] \gets selected\_T_A$
		\State $S[pair[1]] \gets selected\_T_B$
		\EndFor
		\State \textbf{return} $S$
	\end{algorithmic}
\end{algorithm}

%% file: content/evaluation.tex
AWA is independent of the adversary's classifier and generates black-box adversarial traces. To the best of our knowledge, Deep Fingerprinting being proposed by Sirinam \textit{et al.} \cite{DBLP:conf/ccs/SirinamIJW18} is the best adversary's classifier in the previous studies. We use this classifier to evaluate the performance of AWA.
The fundamental assumption of AWA is that although an adversary knows that a transformer set generates the adversarial traces of the target user, she has no knowledge about the details of the target user's transformer set. 
{\color{black}We run AWA several times with various secret random elements to create several transformer sets for evaluating the performance of AWA in this setting.}
We consider two scenarios to evaluate the performance of AWA. 
\begin{enumerate}
	\item
	An adversary randomly selects a set of transformers and trains a classifier on adversarial traces generated by it. 
	The target user also randomly selects a set of transformers and generates her adversarial traces by it. 
	\item
	An adversary generates her adversarial traces through multiple sets of transformers. In this scenario, the adversary must collect more adversarial traces and run a classifier on them, which increases the computational cost of the adversary to run the website fingerprinting attack. The target user also randomly selects a set of transformers and generates her adversarial traces by it. We assume the target user's transformer set is different from the adversary's transformer sets in this scenario. 
\end{enumerate}
Since an adversary has no access to the label of the target user's adversarial traces, she can not evaluate the performance of her classifier on them. Therefore, the adversary's classifier is selected based on the validation set that has been generated by the adversary's transformer set(s). 

\subsection{Dataset and Setup}
We use the dataset that has been proposed by Sirinam \textit{et al.} in \cite{DBLP:conf/ccs/SirinamIJW18}, which consists of traces of 95 different websites, and each website has been visited 1000 times. Each trace of a website is a sequence of packets direction. It is shown in \cite{DBLP:conf/ccs/SirinamIJW18} that sequence of packets direction is enough to classify traces with a high success rate. Because we want the number of classes to be even, we select only 94 classes. The dataset is split into four various sets, including the AWA training set, the adversary's training set, the adversary's validation set, and the target user's set. AWA training set consists of 400 traces of each website (37600 traces in total) and is used in the AWA training to train the auxiliary classifier, generators, and discriminators.  The adversary training and validation sets consist of 400 and 100 traces of each website (37600 and 9400 traces in total) respectively, and are used to train and validate adversaries' classifier. 100 traces of each website (9400 in total) is used as the traces of the target user's browsing activities. 

The architecture of transformers and discriminators are presented in Appendix \ref{sec:appendixa}. Transformers are trained in 1000 iterations where $T_T = 2$, $D_T = 2$, $bs = 100$, and $K=94$. The length of traces being in the format of burst sequence is 2000.
The architecture of the auxiliary classifier is the same as the discriminator. The type of all DNNs is 1D-CNN, and we use Tensorflow 1.12 and Keras 2.2.5 to implement them. Adam optimizer with learning rate 0.0001 is used to optimize parameters of transformers and discriminators. The auxiliary classifier uses Adam optimizer with learning rate 0.0002 and is trained in 30 epochs with batch size 128. The distribution of noise in all experiments is the standard Gaussian distribution. Hyperparameters of the generators' loss function are $\alpha= 10^3, \beta=10^3,$ and $\gamma =10^2$. {\color{black} We run AWA on a single NVIDIA GeForce GTX
	1080 Ti GPU with 11 GB RAM. The training phase of each pair of generators takes about 675 seconds, and generating 100 adversarial traces in the testing phase takes about 18 milliseconds. The computation costs of UAWA and NUAWA are precisely the same.

}

\input{content/perf_tables}

\subsection{Experiments}
\input{content/Experiments}

\subsection{ Intra-Class Distance}
\input{content/old_ClassDistanc}

\subsection{Average Trace Visualization} 
\input{content/visualization}

\subsection{Results Analysis}
\input{content/SimWorks}

%% file: content/perf_tables.tex
\renewcommand{\arraystretch}{1.3}
\newcommand{\matcolor}{red}
\newcommand{\matcoloroh}{yellow}
\setlength\arrayrulewidth{1pt} 

\begin{figure*}[!t]

	\begin{subfigure}{.5\textwidth}
		\centering
		\begin{table}[H]
			\label{tab:my-table}
			\resizebox{1\textwidth}{!}{ 
\begin{tabular}{ccccccccccc}
	\multicolumn{3}{c}{\multirow{2}{*}{Adversary's accuracy(\%)}}                                                                                                   & \multicolumn{5}{c}{Adversary's transformer set}                                                                                                                                                                                                                                                                                                 & \multicolumn{3}{c}{\multirow{3}{*}{\begin{tabular}[c]{@{}c@{}}Combination of all \\ sets but the user's \\ transformer set\end{tabular}}} \\ \cline{4-8}
	\multicolumn{3}{c}{}                                                                                                                                            & Set 1                                                            & Set 2                                                            & Set 3                                                              & Set 4                                                            & Set 5                                                              & \multicolumn{3}{c}{}                                                                                                                      \\ \hhline{~~|-|-|-|-|-|-|}
	\multirow{7}{*}{\rotatebox{90}{User's transformer set}} & \multicolumn{1}{c|}{}      & BWO(\%)                                                                                 & \cellcolor{\matcoloroh!22.4}22.4 & \cellcolor{\matcoloroh!20.9}20.9 & \cellcolor{\matcoloroh!23.46}23.46 & \cellcolor{\matcoloroh!22.0}22.0 & \cellcolor{\matcoloroh!22.29}22.29 & \multicolumn{3}{c}{}                                                                                                                      \\ \hhline{~~|~|-|-|-|-|-|-|-|-|}
	& \multicolumn{1}{|c|}{Set 1} & \multicolumn{1}{c|}{\cellcolor{\matcoloroh!22.43}22.43} & \cellcolor{\matcolor!98.40}98.40 & \cellcolor{\matcolor!18.17}18.17 & \cellcolor{\matcolor!18.76}18.76   & \cellcolor{\matcolor!19.40}19.40 & \cellcolor{\matcolor!19.81}19.81   &        $\quad\;$                 & \cellcolor{\matcolor!63.52}63.52                       &                        \\
	& \multicolumn{1}{|c|}{Set 2} & \multicolumn{1}{c|}{\cellcolor{\matcoloroh!21.03}21.03} & \cellcolor{\matcolor!16.61}16.61 & \cellcolor{\matcolor!98.14}98.14 & \cellcolor{\matcolor!19.59}19.59   & \cellcolor{\matcolor!21.0}21.0   & \cellcolor{\matcolor!20.18}20.18   &                         & \cellcolor{\matcolor!63.68}63.68                       &                        \\
	& \multicolumn{1}{|c|}{Set 3} & \multicolumn{1}{c|}{\cellcolor{\matcoloroh!23.57}23.57} & \cellcolor{\matcolor!17.74}17.74 & \cellcolor{\matcolor!20.67}20.67 & \cellcolor{\matcolor!98.56}98.56   & \cellcolor{\matcolor!22.41}22.41 & \cellcolor{\matcolor!20.09}20.09   &                         & \cellcolor{\matcolor!61.70}61.70                       &                        \\
	& \multicolumn{1}{|c|}{Set 4} & \multicolumn{1}{c|}{\cellcolor{\matcoloroh!22.04}22.04} & \cellcolor{\matcolor!18.05}18.05 & \cellcolor{\matcolor!20.15}20.15 & \cellcolor{\matcolor!19.77}19.77   & \cellcolor{\matcolor!98.11}98.11 & \cellcolor{\matcolor!17.44}17.44   &                         & \cellcolor{\matcolor!64.79}64.79                       &                        \\
	& \multicolumn{1}{|c|}{Set 5} & \multicolumn{1}{c|}{\cellcolor{\matcoloroh!22.35}22.35} & \cellcolor{\matcolor!13.15}13.15 & \cellcolor{\matcolor!23.76}23.76 & \cellcolor{\matcolor!22.71}22.71   & \cellcolor{\matcolor!21.06}21.06 & \cellcolor{\matcolor!98.50}98.50   &                         & \cellcolor{\matcolor!62.68}62.68                       &                        \\
	&                            &                                                                                         &                                                                  &                                                                  &                                                                    &                                                                  &                                                                    &                         &                                                                                        &                       
\end{tabular}
			}
		\end{table}
		 \vspace{-0.5cm}
		\vspace*{-3mm}\caption{\scriptsize UAWA-Exp1: $\tau_{low}=0.05, \tau_{high}=0.30$, and $OH=0.50$}
		\label{tab:UAWA-Exp1}
	\end{subfigure}
\vspace{-.5\baselineskip}
	\begin{subfigure}{.5\textwidth}
		\centering
		\begin{table}[H]

			\label{tab:my-table}
			
			\resizebox{1\textwidth}{!}{ 
				\begin{tabular}{ccccccccccc}
					\multicolumn{3}{c}{\multirow{2}{*}{Adversary's accuracy(\%)}}                                                                                                   & \multicolumn{5}{c}{Adversary's transformer set}                                                                                                                                                                                                                                                                                                     & \multicolumn{3}{c}{\multirow{3}{*}{\begin{tabular}[c]{@{}c@{}}Combination of all \\ sets but the user's \\ transformer set\end{tabular}}} \\ \cline{4-8}
					\multicolumn{3}{c}{}                                                                                                                                            & Set 1                                                              & Set 2                                                              & Set 3                                                              & Set 4                                                            & Set 5                                                              & \multicolumn{3}{c}{}                                                                                                                      \\ \hhline{~~|-|-|-|-|-|-|}
					\multirow{7}{*}{\rotatebox{90}{User's transformer set}} & \multicolumn{1}{c|}{}      & BWO(\%)                                                                                 & \cellcolor{\matcoloroh!26.17}26.17 & \cellcolor{\matcoloroh!26.57}26.57 & \cellcolor{\matcoloroh!26.93}26.93 & \cellcolor{\matcoloroh!26.5}26.5 & \cellcolor{\matcoloroh!26.12}26.12 & \multicolumn{3}{c}{}                                                                                                                      \\ \hhline{~~|~|-|-|-|-|-|-|-|-|}
					& \multicolumn{1}{|c|}{Set 1} & \multicolumn{1}{c|}{\cellcolor{\matcoloroh!25.95}25.95} & \cellcolor{\matcolor!93.53}93.53   & \cellcolor{\matcolor!33.05}33.05   & \cellcolor{\matcolor!27.29}27.29   & \cellcolor{\matcolor!31.01}31.01 & \cellcolor{\matcolor!36.92}36.92   &                         & \cellcolor{\matcolor!61.28}61.28                       &                        \\
					& \multicolumn{1}{|c|}{Set 2} & \multicolumn{1}{c|}{\cellcolor{\matcoloroh!26.29}26.29} & \cellcolor{\matcolor!34.68}34.68   & \cellcolor{\matcolor!94.13}94.13   & \cellcolor{\matcolor!31.72}31.72   & \cellcolor{\matcolor!32.98}32.98 & \cellcolor{\matcolor!34.62}34.62   &           $\quad\;$              & \cellcolor{\matcolor!64.62}64.62                       &                        \\
					& \multicolumn{1}{|c|}{Set 3} & \multicolumn{1}{c|}{\cellcolor{\matcoloroh!27.06}27.06} & \cellcolor{\matcolor!31.0}31.0     & \cellcolor{\matcolor!34.23}34.23   & \cellcolor{\matcolor!93.94}93.94   & \cellcolor{\matcolor!31.12}31.12 & \cellcolor{\matcolor!29.42}29.42   &                         & \cellcolor{\matcolor!58.61}58.61                       &                        \\
					& \multicolumn{1}{|c|}{Set 4} & \multicolumn{1}{c|}{\cellcolor{\matcoloroh!26.15}26.15} & \cellcolor{\matcolor!29.87}29.87   & \cellcolor{\matcolor!32.30}32.30   & \cellcolor{\matcolor!28.63}28.63   & \cellcolor{\matcolor!93.75}93.75 & \cellcolor{\matcolor!32.64}32.64   &                         & \cellcolor{\matcolor!57.82}57.82                       &                        \\
					& \multicolumn{1}{|c|}{Set 5} & \multicolumn{1}{c|}{\cellcolor{\matcoloroh!25.95}25.95} & \cellcolor{\matcolor!35.85}35.85   & \cellcolor{\matcolor!34.36}34.36   & \cellcolor{\matcolor!27.29}27.29   & \cellcolor{\matcolor!30.0}30.0   & \cellcolor{\matcolor!94.01}94.01   &                         & \cellcolor{\matcolor!57.45}57.45                       &                        \\
					&                            &                                                                                         &                                                                    &                                                                    &                                                                    &                                                                  &                                                                    &                         &                                                                                        &                       
				\end{tabular}
			}
		\end{table}
	\vspace{-0.5cm}
		\vspace*{-3mm}\caption{\scriptsize NUAWA-Exp1: $\tau_{low}=0.05, \tau_{high}=0.3 $, and $OH=0.5$}
		\label{tab:NUAWA-Exp1}
	\end{subfigure}
	\vspace{-.5\baselineskip}
	\begin{subfigure}{.5\textwidth}
		\centering
		\begin{table}[H]

			\label{tab:my-table}
			
			\resizebox{1\textwidth}{!}{ 
\begin{tabular}{ccccccccccc}
	\multicolumn{3}{c}{\multirow{2}{*}{Adversary's accuracy(\%)}}                                                                                                   & \multicolumn{5}{c}{Adversary's transformer set}                                                                                                                                                                                                                                                                                                     & \multicolumn{3}{c}{\multirow{3}{*}{\begin{tabular}[c]{@{}c@{}}Combination of all \\ sets but the user's \\ transformer set\end{tabular}}} \\ \cline{4-8}
	\multicolumn{3}{c}{}                                                                                                                                            & Set 1                                                              & Set 2                                                            & Set 3                                                              & Set 4                                                              & Set 5                                                              & \multicolumn{3}{c}{}                                                                                                                      \\ \hhline{~~|-|-|-|-|-|-|}
	\multirow{7}{*}{\rotatebox{90}{User's transformer set}} & \multicolumn{1}{c|}{}      & BWO(\%)                                                                                 & \cellcolor{\matcoloroh!39.15}39.15 & \cellcolor{\matcoloroh!39.3}39.3 & \cellcolor{\matcoloroh!38.47}38.47 & \cellcolor{\matcoloroh!38.65}38.65 & \cellcolor{\matcoloroh!39.67}39.67 & \multicolumn{3}{c}{}                                                                                                                      \\ \hhline{~~|~|-|-|-|-|-|-|-|-|}
	& \multicolumn{1}{|c|}{Set 1} & \multicolumn{1}{c|}{\cellcolor{\matcoloroh!39.28}39.28} & \cellcolor{\matcolor!97.64}97.64   & \cellcolor{\matcolor!6.09}6.09   & \cellcolor{\matcolor!14.73}14.73   & \cellcolor{\matcolor!13.30}13.30   & \cellcolor{\matcolor!18.31}18.31   &               $\quad\;
	$          & \cellcolor{\matcolor!51.17}51.17                       &                        \\
	& \multicolumn{1}{|c|}{Set 2} & \multicolumn{1}{c|}{\cellcolor{\matcoloroh!39.34}39.34} & \cellcolor{\matcolor!7.53}7.53     & \cellcolor{\matcolor!98.51}98.51 & \cellcolor{\matcolor!8.92}8.92     & \cellcolor{\matcolor!8.22}8.22     & \cellcolor{\matcolor!9.6}9.6       &                         & \cellcolor{\matcolor!41.94}41.94                       &                        \\
	& \multicolumn{1}{|c|}{Set 3} & \multicolumn{1}{c|}{\cellcolor{\matcoloroh!38.38}38.38} & \cellcolor{\matcolor!13.58}13.58   & \cellcolor{\matcolor!8.3}8.3     & \cellcolor{\matcolor!97.75}97.75   & \cellcolor{\matcolor!15.8}15.8     & \cellcolor{\matcolor!11.75}11.75   &                         & \cellcolor{\matcolor!54.09}54.09                       &                        \\
	& \multicolumn{1}{|c|}{Set 4} & \multicolumn{1}{c|}{\cellcolor{\matcoloroh!38.65}38.65} & \cellcolor{\matcolor!13.32}13.32   & \cellcolor{\matcolor!7.57}7.57   & \cellcolor{\matcolor!11.44}11.44   & \cellcolor{\matcolor!98.42}98.42   & \cellcolor{\matcolor!9.57}9.57     &                         & \cellcolor{\matcolor!42.48}42.48                       &                        \\
	& \multicolumn{1}{|c|}{Set 5} & \multicolumn{1}{c|}{\cellcolor{\matcoloroh!39.65}39.65} & \cellcolor{\matcolor!18.08}18.08   & \cellcolor{\matcolor!7.89}7.89   & \cellcolor{\matcolor!11.03}11.03   & \cellcolor{\matcolor!12.95}12.95   & \cellcolor{\matcolor!97.71}97.71   &                         & \cellcolor{\matcolor!52.82}52.82                       &                        \\
	&                            &                                                                                         &                                                                    &                                                                  &                                                                    &                                                                    &                                                                    &                         &                                                                                        &                       
\end{tabular}
			}
		\end{table} 
	\vspace{-0.5cm}
		\vspace*{-3mm}\caption{\scriptsize UAWA-Exp2: $\tau_{low}=0.25, \tau_{high}=0.5$, and $OH=0.75$}
		\label{tab:UAWA-Exp2}
	\end{subfigure}
	\begin{subfigure}{.5\textwidth}
		\centering
		
		\begin{table}[H]

			\label{tab:my-table}
			
			\resizebox{1\textwidth}{!}{ 
\begin{tabular}{ccccccccccc}
	\multicolumn{3}{c}{\multirow{2}{*}{Adversary's accuracy(\%)}}                                                                                                   & \multicolumn{5}{c}{Adversary's transformer set}                                                                                                                                                                                                                                                                                                   & \multicolumn{3}{c}{\multirow{3}{*}{\begin{tabular}[c]{@{}c@{}}Combination of all \\ sets but the user's \\ transformer set\end{tabular}}} \\ \cline{4-8}
	\multicolumn{3}{c}{}                                                                                                                                            & Set 1                                                            & Set 2                                                              & Set 3                                                              & Set 4                                                            & Set 5                                                              & \multicolumn{3}{c}{}                                                                                                                      \\ \hhline{~~|-|-|-|-|-|-|}
	\multirow{7}{*}{\rotatebox{90}{User's transformer set}} & \multicolumn{1}{c|}{}      & BWO(\%)                                                                                 & \cellcolor{\matcoloroh!43.3}43.3 & \cellcolor{\matcoloroh!42.17}42.17 & \cellcolor{\matcoloroh!41.81}41.81 & \cellcolor{\matcoloroh!42.0}42.0 & \cellcolor{\matcoloroh!43.38}43.38 & \multicolumn{3}{c}{}                                                                                                                      \\ \hhline{~~|~|-|-|-|-|-|-|-|-|}
	& \multicolumn{1}{|c|}{Set 1} & \multicolumn{1}{c|}{\cellcolor{\matcoloroh!43.4}43.4}   & \cellcolor{\matcolor!95.47}95.47 & \cellcolor{\matcolor!19.63}19.63   & \cellcolor{\matcolor!17.29}17.29   & \cellcolor{\matcolor!19.38}19.38 & \cellcolor{\matcolor!18.93}18.93   & $\quad\;$                     & \cellcolor{\matcolor!42.89}42.89                    &                     \\
	& \multicolumn{1}{|c|}{Set 2} & \multicolumn{1}{c|}{\cellcolor{\matcoloroh!41.92}41.92} & \cellcolor{\matcolor!21.41}21.41 & \cellcolor{\matcolor!95.61}95.61   & \cellcolor{\matcolor!16.47}16.47   & \cellcolor{\matcolor!16.63}16.63 & \cellcolor{\matcolor!19.06}19.06   &                               & \cellcolor{\matcolor!42.13}42.13                    &                     \\
	& \multicolumn{1}{|c|}{Set 3} & \multicolumn{1}{c|}{\cellcolor{\matcoloroh!41.67}41.67} & \cellcolor{\matcolor!20.86}20.86 & \cellcolor{\matcolor!18.85}18.85   & \cellcolor{\matcolor!93.52}93.52   & \cellcolor{\matcolor!17.09}17.09 & \cellcolor{\matcolor!19.30}19.30   &                               & \cellcolor{\matcolor!39.54}39.54                    &                     \\
	& \multicolumn{1}{|c|}{Set 4} & \multicolumn{1}{c|}{\cellcolor{\matcoloroh!41.97}41.97} & \cellcolor{\matcolor!19.17}19.17 & \cellcolor{\matcolor!20.02}20.02   & \cellcolor{\matcolor!14.52}14.52   & \cellcolor{\matcolor!95.37}95.37 & \cellcolor{\matcolor!15.85}15.85   &                               & \cellcolor{\matcolor!39.70}39.70                    &                     \\
	& \multicolumn{1}{|c|}{Set 5} & \multicolumn{1}{c|}{\cellcolor{\matcoloroh!43.53}43.53} & \cellcolor{\matcolor!18.80}18.80 & \cellcolor{\matcolor!19.65}19.65   & \cellcolor{\matcolor!17.82}17.82   & \cellcolor{\matcolor!15.18}15.18 & \cellcolor{\matcolor!94.48}94.48   &                               & \cellcolor{\matcolor!39.48}39.48                    &                     \\
	&                            &                                                                                         &                                                                  &                                                                    &                                                                    &                                                                  &                                                                    &                               &                                                                                     &                    
\end{tabular}
			}
		\end{table}  
	\vspace{-0.5cm}
		\vspace*{-3mm}\caption{\scriptsize NUAWA-Exp2: $\tau_{low}=0.25, \tau_{high}=0.5$, and $OH=0.75$}
		\label{tab:NUAWA-Exp2}
	\end{subfigure}
\vspace{-.5\baselineskip}
	\begin{subfigure}{.5\textwidth}
	\centering
\begin{table}[H]

	\label{tab:my-table}
	
	\resizebox{1\textwidth}{!}{ 
\begin{tabular}{ccccccccccc}
	\multicolumn{3}{c}{\multirow{2}{*}{Adversary's accuracy(\%)}}                                                                                                   & \multicolumn{5}{c}{Adversary's transformer set}                                                                                                                                                                                                                                                                                                   & \multicolumn{3}{c}{\multirow{3}{*}{\begin{tabular}[c]{@{}c@{}}Combination of all \\ sets but the user's \\ transformer set\end{tabular}}} \\ \cline{4-8}
	\multicolumn{3}{c}{}                                                                                                                                            & Set 1                                                            & Set 2                                                              & Set 3                                                            & Set 4                                                              & Set 5                                                              & \multicolumn{3}{c}{}                                                                                                                      \\ \hhline{~~|-|-|-|-|-|-|}
	\multirow{7}{*}{\rotatebox{90}{User's transformer set}} & \multicolumn{1}{c|}{}      & BWO(\%)                                                                                 & \cellcolor{\matcoloroh!62.6}62.6 & \cellcolor{\matcoloroh!61.78}61.78 & \cellcolor{\matcoloroh!61.1}61.1 & \cellcolor{\matcoloroh!63.89}63.89 & \cellcolor{\matcoloroh!63.54}63.54 & \multicolumn{3}{c}{}                                                                                                                      \\ \hhline{~~|~|-|-|-|-|-|-|-|-|}
	& \multicolumn{1}{|c|}{Set 1} & \multicolumn{1}{c|}{\cellcolor{\matcoloroh!62.61}62.61} & \cellcolor{\matcolor!98.30}98.30 & \cellcolor{\matcolor!6.70}6.70     & \cellcolor{\matcolor!8.58}8.58   & \cellcolor{\matcolor!7.58}7.58     & \cellcolor{\matcolor!8.58}8.58     & $\quad\;$                     & \cellcolor{\matcolor!51.65}51.65                    &                     \\
	& \multicolumn{1}{|c|}{Set 2} & \multicolumn{1}{c|}{\cellcolor{\matcoloroh!61.68}61.68} & \cellcolor{\matcolor!7.62}7.62   & \cellcolor{\matcolor!98.96}98.96   & \cellcolor{\matcolor!8.30}8.30   & \cellcolor{\matcolor!7.67}7.67     & \cellcolor{\matcolor!13.42}13.42   &                               & \cellcolor{\matcolor!51.09}51.09                    &                     \\
	& \multicolumn{1}{|c|}{Set 3} & \multicolumn{1}{c|}{\cellcolor{\matcoloroh!61.07}61.07} & \cellcolor{\matcolor!4.85}4.85   & \cellcolor{\matcolor!6.74}6.74     & \cellcolor{\matcolor!98.77}98.77 & \cellcolor{\matcolor!8.27}8.27     & \cellcolor{\matcolor!11.93}11.93   &                               & \cellcolor{\matcolor!49.10}49.10                    &                     \\
	& \multicolumn{1}{|c|}{Set 4} & \multicolumn{1}{c|}{\cellcolor{\matcoloroh!63.73}63.73} & \cellcolor{\matcolor!10.20}10.20 & \cellcolor{\matcolor!5.20}5.20     & \cellcolor{\matcolor!11.61}11.61 & \cellcolor{\matcolor!99.13}99.13   & \cellcolor{\matcolor!8.13}8.13     &                               & \cellcolor{\matcolor!43.85}43.85                    &                     \\
	& \multicolumn{1}{|c|}{Set 5} & \multicolumn{1}{c|}{\cellcolor{\matcoloroh!63.52}63.52} & \cellcolor{\matcolor!8.55}8.55   & \cellcolor{\matcolor!11.85}11.85   & \cellcolor{\matcolor!10.37}10.37 & \cellcolor{\matcolor!7.76}7.76     & \cellcolor{\matcolor!98.69}98.69   &                               & \cellcolor{\matcolor!49.84}49.84                    &                     \\
	&                            &                                                                                         &                                                                  &                                                                    &                                                                  &                                                                    &                                                                    &                               &                                                                                     &                    
\end{tabular}
	}
\end{table}
\vspace{-0.5cm}
	\vspace*{-3mm}\caption{\scriptsize UAWA-Exp3: $\tau_{low}=0.5, \tau_{high}=0.75$, and $OH=1.0$}
	\label{tab:UAWA-Exp3}
\end{subfigure}
\begin{subfigure}{.5\textwidth}
	\centering
	
\begin{table}[H]

	\label{tab:my-table}
	
	\resizebox{1\textwidth}{!}{ 
		\begin{tabular}{ccccccccccc}
			\multicolumn{3}{c}{\multirow{2}{*}{Adversary's accuracy(\%)}}    & \multicolumn{5}{c}{Adversary's transformer set}                                                                                                                                                                                                                                                                                                       & \multicolumn{3}{c}{\multirow{3}{*}{\begin{tabular}[c]{@{}c@{}}Combination of all \\ sets but the user's \\ transformer set\end{tabular}}} \\ \cline{4-8}
			&                            &                                                                                         & Set 1                                                              & Set 2                                                              & Set 3                                                              & Set 4                                                              & Set 5                                                              & \multicolumn{3}{c}{}                                                                                                                      \\ \hhline{~~|-|-|-|-|-|-|}
			\multirow{7}{*}{\rotatebox{90}{User's transformer set}} & \multicolumn{1}{c|}{}      & BWO(\%)                                                                                 & \cellcolor{\matcoloroh!64.43}64.43 & \cellcolor{\matcoloroh!65.53}65.53 & \cellcolor{\matcoloroh!63.44}63.44 & \cellcolor{\matcoloroh!63.77}63.77 & \cellcolor{\matcoloroh!64.95}64.95 & \multicolumn{3}{c}{}                                                                                                                      \\ \hhline{~~|~|-|-|-|-|-|-|-|-|}
			& \multicolumn{1}{|c|}{Set 1} & \multicolumn{1}{c|}{\cellcolor{\matcoloroh!64.28}64.28} & \cellcolor{\matcolor!95.57}95.57   & \cellcolor{\matcolor!11.69}11.69   & \cellcolor{\matcolor!12.82}12.82   & \cellcolor{\matcolor!13.22}13.22   & \cellcolor{\matcolor!9.53}9.53     &             $\quad\;$            & \cellcolor{\matcolor!27.80}27.80                       &                        \\
			& \multicolumn{1}{|c|}{Set 2} & \multicolumn{1}{c|}{\cellcolor{\matcoloroh!65.15}65.15} & \cellcolor{\matcolor!11.54}11.54   & \cellcolor{\matcolor!96.58}96.58   & \cellcolor{\matcolor!13.54}13.54   & \cellcolor{\matcolor!11.57}11.57   & \cellcolor{\matcolor!11.76}11.76   &                         & \cellcolor{\matcolor!26.22}26.22                       &                        \\
			& \multicolumn{1}{|c|}{Set 3} & \multicolumn{1}{c|}{\cellcolor{\matcoloroh!63.74}63.74} & \cellcolor{\matcolor!14.87}14.87   & \cellcolor{\matcolor!13.11}13.11   & \cellcolor{\matcolor!97.17}97.17   & \cellcolor{\matcolor!9.00}9.00     & \cellcolor{\matcolor!12.96}12.96   &                         & \cellcolor{\matcolor!26.97}26.97                       &                        \\
			& \multicolumn{1}{|c|}{Set 4} & \multicolumn{1}{c|}{\cellcolor{\matcoloroh!63.97}63.97} & \cellcolor{\matcolor!12.73}12.73   & \cellcolor{\matcolor!11.06}11.06   & \cellcolor{\matcolor!10.09}10.09   & \cellcolor{\matcolor!96.12}96.12   & \cellcolor{\matcolor!15.11}15.11   &                         & \cellcolor{\matcolor!23.63}23.63                       &                        \\
			& \multicolumn{1}{|c|}{Set 5} & \multicolumn{1}{c|}{\cellcolor{\matcoloroh!64.54}64.54} & \cellcolor{\matcolor!9.89}9.89     & \cellcolor{\matcolor!9.64}9.64     & \cellcolor{\matcolor!13.25}13.25   & \cellcolor{\matcolor!13.87}13.87   & \cellcolor{\matcolor!95.93}95.93   &                         & \cellcolor{\matcolor!25.05}25.05                       &                        \\
			&                            &                                                                                         &                                                                    &                                                                    &                                                                    &                                                                    &                                                                    &                         &                                                                                        &                       
		\end{tabular}
	}
\end{table}

\vspace{-0.5cm}
	\vspace*{-3mm}\caption{\scriptsize NUAWA-Exp3: $\tau_{low}=0.5, \tau_{high}=0.75$, and $OH=1.00$}
	\label{tab:NUAWA-Exp3}
\end{subfigure}

\begin{subfigure}{.5\textwidth}
	\centering
	\begin{table}[H]
		
		\label{tab:my-table}
		
		\resizebox{1\textwidth}{!}{ 
			\begin{tabular}{ccccccccccc}
				\multicolumn{3}{c}{\multirow{2}{*}{Adversary's accuracy(\%)}}                                                                                                   & \multicolumn{5}{c}{Adversary's transformer set}                                                                                                                                                                                                                                                                                                     & \multicolumn{3}{c}{\multirow{3}{*}{\begin{tabular}[c]{@{}c@{}}Combination of all \\ sets but the user's \\ transformer set\end{tabular}}} \\ \cline{4-8}
				\multicolumn{3}{c}{}                                                                                                                                            & Set 1                                                              & Set 2                                                              & Set 3                                                              & Set 4                                                            & Set 5                                                              & \multicolumn{3}{c}{}                                                                                                                      \\ \hhline{~~|-|-|-|-|-|-|}
				\multirow{7}{*}{\rotatebox{90}{User's transformer set}} & \multicolumn{1}{c|}{}      & BWO(\%)                                                                                 & \cellcolor{\matcoloroh!86.13}86.13 & \cellcolor{\matcoloroh!87.24}87.24 & \cellcolor{\matcoloroh!84.92}84.92 & \cellcolor{\matcoloroh!85.3}85.3 & \cellcolor{\matcoloroh!86.85}86.85 & \multicolumn{3}{c}{}                                                                                                                      \\ \hhline{~~|~|-|-|-|-|-|-|-|-|}
				& \multicolumn{1}{|c|}{Set 1} & \multicolumn{1}{c|}{\cellcolor{\matcoloroh!86.25}86.25} & \cellcolor{\matcolor!99.03}99.03   & \cellcolor{\matcolor!6.44}6.44     & \cellcolor{\matcolor!9.87}9.87     & \cellcolor{\matcolor!6.26}6.26   & \cellcolor{\matcolor!2.28}2.28     & $\quad\;$                     & \cellcolor{\matcolor!52.02}52.02                    &                     \\
				& \multicolumn{1}{|c|}{Set 2} & \multicolumn{1}{c|}{\cellcolor{\matcoloroh!87.15}87.15} & \cellcolor{\matcolor!8.45}8.45     & \cellcolor{\matcolor!98.93}98.93   & \cellcolor{\matcolor!4.42}4.42     & \cellcolor{\matcolor!6.34}6.34   & \cellcolor{\matcolor!5.51}5.51     &                               & \cellcolor{\matcolor!52.34}52.34                    &                     \\
				& \multicolumn{1}{|c|}{Set 3} & \multicolumn{1}{c|}{\cellcolor{\matcoloroh!84.88}84.88} & \cellcolor{\matcolor!9.97}9.97     & \cellcolor{\matcolor!6.24}6.24     & \cellcolor{\matcolor!98.28}98.28   & \cellcolor{\matcolor!7.03}7.03   & \cellcolor{\matcolor!9.95}9.95     &                               & \cellcolor{\matcolor!50.14}50.14                    &                     \\
				& \multicolumn{1}{|c|}{Set 4} & \multicolumn{1}{c|}{\cellcolor{\matcoloroh!85.45}85.45} & \cellcolor{\matcolor!6.75}6.75     & \cellcolor{\matcolor!6.10}6.10     & \cellcolor{\matcolor!4.47}4.47     & \cellcolor{\matcolor!98.73}98.73 & \cellcolor{\matcolor!8.23}8.23     &                               & \cellcolor{\matcolor!39.88}39.88                    &                     \\
				& \multicolumn{1}{|c|}{Set 5} & \multicolumn{1}{c|}{\cellcolor{\matcoloroh!86.77}86.77} & \cellcolor{\matcolor!7.17}7.17     & \cellcolor{\matcolor!6.62}6.62     & \cellcolor{\matcolor!10.63}10.63   & \cellcolor{\matcolor!9.17}9.17   & \cellcolor{\matcolor!98.98}98.98   &                               & \cellcolor{\matcolor!47.46}47.46                    &                     \\
				&                            &                                                                                         &                                                                    &                                                                    &                                                                    &                                                                  &                                                                    &                               &                                                                                     &                    
			\end{tabular}
		}
	\end{table}
	\vspace{-0.5cm}
	\vspace*{-3mm}\caption{\scriptsize UAWA-Exp4: $\tau_{low}=0.75, \tau_{high}=1.0$, and $OH=1.25$}
	\label{tab:UAWA-Exp4}
\end{subfigure}
\begin{subfigure}{.5\textwidth}
	\centering
	
	\begin{table}[H]
		
		\label{tab:my-table}
		
		\resizebox{1\textwidth}{!}{ 
\begin{tabular}{ccccccccccc}
	\multicolumn{3}{c}{\multirow{2}{*}{Adversary's accuracy(\%)}}                                                                                                   & \multicolumn{5}{c}{Adversary's transformer set}                                                                                                                                                                                                                                                                                                       & \multicolumn{3}{c}{\multirow{3}{*}{\begin{tabular}[c]{@{}c@{}}Combination of all \\ sets but the user's \\ transformer set\end{tabular}}} \\ \cline{4-8}
	\multicolumn{3}{c}{}                                                                                                                                            & Set 1                                                              & Set 2                                                              & Set 3                                                              & Set 4                                                              & Set 5                                                              & \multicolumn{3}{c}{}                                                                                                                      \\ \hhline{~~|-|-|-|-|-|-|}
	\multirow{7}{*}{\rotatebox{90}{User's transformer set}} & \multicolumn{1}{c|}{}      & BWO(\%)                                                                                 & \cellcolor{\matcoloroh!87.59}87.59 & \cellcolor{\matcoloroh!87.09}87.09 & \cellcolor{\matcoloroh!86.76}86.76 & \cellcolor{\matcoloroh!86.23}86.23 & \cellcolor{\matcoloroh!89.67}89.67 & \multicolumn{3}{c}{}                                                                                                                      \\ \hhline{~~|~|-|-|-|-|-|-|-|-|}
	& \multicolumn{1}{|c|}{Set 1} & \multicolumn{1}{c|}{\cellcolor{\matcoloroh!88.08}88.08} & \cellcolor{\matcolor!97.41}97.41   & \cellcolor{\matcolor!8.91}8.91     & \cellcolor{\matcolor!10.55}10.55   & \cellcolor{\matcolor!7.44}7.44     & \cellcolor{\matcolor!9.04}9.04     & $\quad\;$                     & \cellcolor{\matcolor!21.32}21.32                    &                     \\
	& \multicolumn{1}{|c|}{Set 2} & \multicolumn{1}{c|}{\cellcolor{\matcoloroh!86.72}86.72} & \cellcolor{\matcolor!6.70}6.70     & \cellcolor{\matcolor!97.59}97.59   & \cellcolor{\matcolor!6.95}6.95     & \cellcolor{\matcolor!9.81}9.81     & \cellcolor{\matcolor!10.40}10.40   &                               & \cellcolor{\matcolor!23.26}23.26                    &                     \\
	& \multicolumn{1}{|c|}{Set 3} & \multicolumn{1}{c|}{\cellcolor{\matcoloroh!86.89}86.89} & \cellcolor{\matcolor!12.08}12.08   & \cellcolor{\matcolor!10.64}10.64   & \cellcolor{\matcolor!97.17}97.17   & \cellcolor{\matcolor!8.57}8.57     & \cellcolor{\matcolor!7.17}7.17     &                               & \cellcolor{\matcolor!21.84}21.84                    &                     \\
	& \multicolumn{1}{|c|}{Set 4} & \multicolumn{1}{c|}{\cellcolor{\matcoloroh!86.32}86.32} & \cellcolor{\matcolor!9.26}9.26     & \cellcolor{\matcolor!7.18}7.18     & \cellcolor{\matcolor!7.35}7.35     & \cellcolor{\matcolor!97.26}97.26   & \cellcolor{\matcolor!12.77}12.77   &                               & \cellcolor{\matcolor!22.97}22.97                    &                     \\
	& \multicolumn{1}{|c|}{Set 5} & \multicolumn{1}{c|}{\cellcolor{\matcoloroh!88.97}88.97} & \cellcolor{\matcolor!8.52}8.52     & \cellcolor{\matcolor!9.13}9.13     & \cellcolor{\matcolor!5.63}5.63     & \cellcolor{\matcolor!10.21}10.21   & \cellcolor{\matcolor!97.95}97.95   &                               & \cellcolor{\matcolor!20.44}20.44                    &                     \\
	&                            &                                                                                         &                                                                    &                                                                    &                                                                    &                                                                    &                                                                    &                               &                                                                                     &                    
\end{tabular}
		}
	\end{table}

	\vspace{-0.5cm}
	\vspace*{-3mm}\caption{\scriptsize NUAWA-Exp4: $\tau_{low}=0.75, \tau_{high}=1.0$, and $OH=1.25$}
	\label{tab:NUAWA-Exp4}
\end{subfigure}

	\caption{ \footnotesize Each table shows the performance of an experiment. The tables on the left and the right sides of the figure indicate the performance of UAWA and NUAWA, respectively, in various magnitudes of Bandwidth Overhead (BWO). Each table shows the accuracy of the adversary's classifiers in all possible ways that an adversary and a target user can select a set of transformers. Each table also indicates the magnitude of BWO that is imposed on the traces of an adversary and a target user. The last column of each table shows the accuracy of the adversary's classifier when an adversary generates her traces through four various sets of transformers, in which the transformer set of the target user is not included.}
	\label{fig:exps_perf}

\end{figure*}

%% file: content/Experiments.tex
We conduct eight experiments to evaluate the performance of AWA. 
We run the AWA training phase for five times with various secret random elements and create five sets of transformers for each experiment. Each table in Figure \ref{fig:exps_perf} shows the adversary's classifier accuracy on all possible ways that an adversary and a target user can pick a set of transformers. Each table also shows the magnitude of BWO that a set of transformers impose on the adversary's traces and the target user's traces. The last column of each table indicates the adversary's classifier accuracy in scenario 2. In this setting, the adversary's training set consists of 400$\times$4$\times$94= 150,400 adversarial traces.
The parameters $\tau_{low}, \tau_{high}$ and $OH$ control the magnitude of BandWidth Overhead (BWO) in experiments and are called BWO controlling parameters.
Figure \ref{fig:exps_perf} shows the results of each experiment in a single table. Four tables on the left side of Figure \ref{fig:exps_perf} indicate the performance of UAWA under the various BWO controlling parameters, and the other four experiments on the right side of Figure \ref{fig:exps_perf} show the performance of NUAWA under the various BWO controlling parameters. The tables that are next to each other have the same BWO controlling parameters, and BWO is increasing from the top to the bottom in each column.

 \input{content/sumperfexpsplot}
\input{content/ohplot}
\input{content/plots}

The first experiment of UAWA (UAWA-Exp1) which results are shown in Table \ref{tab:UAWA-Exp1} demonstrates that UAWA is very effective against website fingerprinting attack, and if the transformer sets of an adversary and a target user are different, then the adversary's classifier accuracy is almost 19.52\% with almost 22.28\% BWO. The last column of Table \ref{tab:UAWA-Exp1} indicates that the accuracy of adversary's classifier in scenario 2 is almost 63.27\%. 
BWO is increased by changing BWO controlling parameters in the second, third, and fourth experiments of UAWA (UAWA-Exp2, UAWA-Exp3, and UAWA-Exp4), which their results are shown in Table \ref{tab:UAWA-Exp2}, Table \ref{tab:UAWA-Exp3}, and Table \ref{tab:UAWA-Exp4}, respectively.
The results of these four experiments demonstrate that UAWA is very effective against the website fingerprinting attack in scenario 1, and the accuracy of adversary's classifier is more decreased by adding more BWO. 
However, when an adversary is more powerful and can generate adversarial traces through multiple sets of transformers, UAWA is not very effective, and the accuracy of adversary's classifier is not decreased too much by increasing BWO between UAWA-Exp1 and UAWA-Exp2, and it is relatively fixed between UAWA-Exp2, UAWA-Exp3, and UAWA-Exp4.

We replicate four experiments of UAWA with the same parameters for UNAWA. 
The results of NUAWA-Exp1, NUAWA-Exp2, NUAWA-Exp3, and NUAWA-Exp4 are presented in Tables \ref{tab:NUAWA-Exp1}, \ref{tab:NUAWA-Exp2}, \ref{tab:NUAWA-Exp3}, and \ref{tab:NUAWA-Exp4}, respectively.
The results of these four experiments of NUAWA demonstrate that NUAWA is effective against website fingerprinting attack, and the accuracy of the adversary's classifier is more decreased by adding more BWO to traces. Although NUAWA is less effective than UAWA in scenario 1, NUAWA is considerably more effective than UAWA in scenario 2. For example in a comparison between UAWA-Exp3 and NUAWA-Exp3 which their BWO is almost the same, although the adversary's classifier accuracy in UAWA is almost 3.3\% less than NUAWA in scenario 1, the adversary's classifier accuracy in NUAWA is almost 23\% less than UAWA in scenario 2.
The results of all experiments demonstrate that if the transformer sets of an adversary and a target user are the same, the accuracy of adversary's classifier is high. However, an adversary can not evaluate her classifier on the target user adversarial traces because she has no access to their labels. Therefore, an adversary can not determine whether the transformer sets are the same, and she has no trust in her classifier accuracy.

Figure \ref{fig:accBWOtrend} summarizes the relationship between adversary's classifier accuracy and BWO in all eight experiments in scenarios 1 and 2. 
{\color{black}
The results demonstrate that the standard deviation between the adversary's classifier accuracy in various experiments is low. Hence, transformer sets that are generated by various random elements have similar performance.}
Figure \ref{fig:BWOClass} shows the average BWO of each class for four various experiments. The results demonstrate that BWO controlling parameters are very effective in controlling BWO, and BWO of almost all classes are in the range [$\tau_{low}$,$\tau_{high}$].

%% file: content/sumperfexpsplot.tex
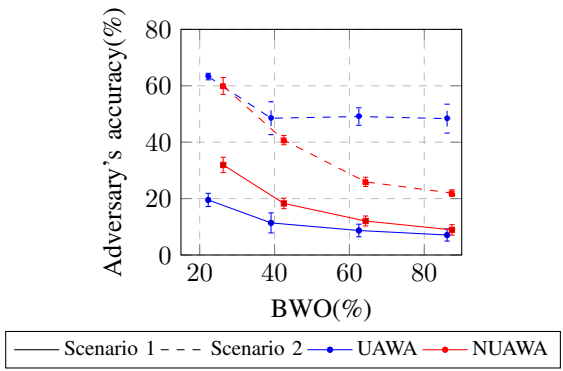
\begin{figure}[]
	\centering
	\begin{subfigure}{0.8\linewidth}
		\centering
		\begin{tikzpicture}
		\begin{axis}[
		xmin=15,   xmax=90,
		ymin=0,   ymax=80,
		xlabel=BWO(\%),
		ylabel=Adversary's accuracy(\%),
		grid=major,
		width=0.75\textwidth,
		]
		\addplot+[,color=blue,mark=*,blue,mark size=1pt,error bars/.cd,y dir=both,y explicit] 
		coordinates {
			(22.28,19.52)   +- (2.34,2.34)
			(39.06,11.39) +- (3.55,3.55)
			(62.52,8.69)   +- (2.24,2.24)
			(86.1,7.09)   +- (2.14,2.14)
		};
		\addplot+[,color=red,mark=square*,red,mark size=1pt,error bars/.cd,y dir=both,y explicit] 
		coordinates {
			(26.28,31.94)   +- (2.71,2.71)
			(42.49,18.29) +- (1.86,1.86)
			(64.33,12.06)   +- (1.78,1.78)
			(87.39,8.91)   +- (1.86,1.86)
		};
		\addplot+[,color=blue,dashed,mark=*,blue,mark size=1pt,mark options={fill=blue},error bars/.cd,y dir=both,y explicit] 
		coordinates {
			(22.28,63.27)   +- (1.15,1.15)
			(39.06,48.5) +- (5.83,5.83)
			(62.52,49.10)   +- (3.10,3.10)
			(86.1,48.36)   +- (5.12,5.12)
		};
		\addplot+[,color=red,dashed,mark=square*,red,mark size=1pt,error bars/.cd,y dir=both,y explicit] 
		coordinates {
			(26.28,59.95)   +- (3.0,3.0)
			(42.49,40.74) +- (1.63,1.63)
			(64.33,25.93)   +- (1.63,1.63)
			(87.39,21.96)   +- (1.16,1.16)
		};
		\end{axis}
		\end{tikzpicture}
		\label{fig:fcc_evala}
	\end{subfigure}
	
	\begin{subfigure}{\linewidth}
		\centering
		\begin{tikzpicture} 
		\begin{axis}[%
		hide axis,
		xmin=10,
		xmax=50,
		ymin=0,
		ymax=0.4,
		legend style={nodes={scale=0.8, transform shape},at={($(0,0)+(1cm,1cm)$)},legend columns=-1,fill=none,draw=black,anchor=center,align=center}
		]
		\addlegendimage{black}
		\addlegendentry{Scenario 1};  
		\addlegendimage{black,dashed}
		\addlegendentry{Scenario 2};   
		\addlegendimage{blue,mark=*,mark size=1pt,mark options={fill=blue}}
		\addlegendentry{UAWA};  
		\addlegendimage{red,mark=*,mark size=1pt,mark options={fill=red}}
		\addlegendentry{NUAWA};

		\end{axis}
		\end{tikzpicture}
	\end{subfigure}
	\caption{\footnotesize The average and standard deviation of the adversary's classifier accuracy over various sizes of BWO in both scenarios.}
	\label{fig:accBWOtrend}
\end{figure}

%% file: content/ohplot.tex
\begin{figure*}[!t]
	\centering
	\begin{tikzpicture}
	\pgfplotsset{
		ybar,  
		scaled y ticks = false,
		width=5.5cm,
		height=3.5cm,
		axis on top,
		xmin=-2,xmax=95,
		ymin=0,ymax=100,
		minor y tick num=1,
		label style = {font = {\fontsize{8 pt}{12 pt}\selectfont}},
		title style = {font = {\fontsize{9pt}{12 pt}\selectfont}},
		tick label style = {font = {\fontsize{4 pt}{12 pt}\selectfont}},
		xtick={-5},
		xticklabels={5},
		y tick label style={rotate=90},
		ylabel shift={-.3em},
		ylabel style={align=center},
		xlabel shift={-0.4em},
	}
	\begin{groupplot}[
	group style={
		group name=my plots,
		group size=4 by 1,
		vertical sep=40pt,
		horizontal sep=15pt},
	legend style={legend columns=-1,at={(0.8,0.17)},
		anchor=north}
	]
	\nextgroupplot[ylabel={BWO(\%) / 100},bar width=0.001,
	ytick={5, 30},yticklabels={\tiny \textcolor{red}{$\tau_{low}$}, \tiny \textcolor{red}{$\tau_{high}$}},]
	\addplot coordinates {(0,26.018)
		(1,17.114)
		(2,17.342)
		(3,14.644)
		(4,16.784)
		(5,13.126)
		(6,20.718)
		(7,22.614)
		(8,26.146)
		(9,15.598)
		(10,19.888)
		(11,27.402)
		(12,24.332)
		(13,17.246)
		(14,12.074)
		(15,18.928)
		(16,25.704)
		(17,23.754)
		(18,26.34)
		(19,20.25)
		(20,27.608)
		(21,28.798)
		(22,8.848)
		(23,21.22)
		(24,28.114)
		(25,12.572)
		(26,16.898)
		(27,23.274)
		(28,19.31)
		(29,20.388)
		(30,13.452)
		(31,18.26)
		(32,14.416)
		(33,29.402)
		(34,24.242)
		(35,22.598)
		(36,19.32)
		(37,16.058)
		(38,19.77)
		(39,23.812)
		(40,20.6)
		(41,19.184)
		(42,14.968)
		(43,28.346)
		(44,24.354)
		(45,28.342)
		(46,19.354)
		(47,31.616)
		(48,13.162)
		(49,26.442)
		(50,28.284)
		(51,20.108)
		(52,25.292)
		(53,16.626)
		(54,28.688)
		(55,21.87)
		(56,24.778)
		(57,17.59)
		(58,21.538)
		(59,18.732)
		(60,24.314)
		(61,26.998)
		(62,28.952)
		(63,24.274)
		(64,14.914)
		(65,16.244)
		(66,15.106)
		(67,15.326)
		(68,21.626)
		(69,27.428)
		(70,21.882)
		(71,15.276)
		(72,23.592)
		(73,21.29)
		(74,9.28)
		(75,24.166)
		(76,23.062)
		(77,18.404)
		(78,24.588)
		(79,20.768)
		(80,19.374)
		(81,26.638)
		(82,27.59)
		(83,24.912)
		(84,18.828)
		(85,24.538)
		(86,23.788)
		(87,18.788)
		(88,14.2)
		(89,21.78)
		(90,22.712)
		(91,26.86)
		(92,21.158)
		(93,22.366)
	};  
	\draw [red] ({rel axis cs:0,0}|-{axis cs:93,5}) -- ({rel axis cs:1,0}|-{axis cs:93,5}) ;
	\draw [red] ({rel axis cs:0,0}|-{axis cs:93,30}) -- ({rel axis cs:1,0}|-{axis cs:93,30}) ;
	
	\nextgroupplot[bar width=0.001,
	ytick={5, 30},yticklabels={\tiny \textcolor{red}{$\tau_{low}$}, \tiny \textcolor{red}{$\tau_{high}$}},]
	\addplot coordinates {(0,28.69)
		(1,24.01)
		(2,30.824)
		(3,22.396)
		(4,25.002)
		(5,25.18)
		(6,29.312)
		(7,21.026)
		(8,28.43)
		(9,24.254)
		(10,27.686)
		(11,29.346)
		(12,28.124)
		(13,23.098)
		(14,28.576)
		(15,29.732)
		(16,27.686)
		(17,24.706)
		(18,27.036)
		(19,22.434)
		(20,26.036)
		(21,26.846)
		(22,18.44)
		(23,24.53)
		(24,29.28)
		(25,25.43)
		(26,23.908)
		(27,26.034)
		(28,22.386)
		(29,25.142)
		(30,31.222)
		(31,20.948)
		(32,25.854)
		(33,28.728)
		(34,29.73)
		(35,27.76)
		(36,27.24)
		(37,30.082)
		(38,26.758)
		(39,22.65)
		(40,28.298)
		(41,28.634)
		(42,22.69)
		(43,32.158)
		(44,28.476)
		(45,26.632)
		(46,28.832)
		(47,28.802)
		(48,25.306)
		(49,29.906)
		(50,24.602)
		(51,21.35)
		(52,23.676)
		(53,22.836)
		(54,23.922)
		(55,33.798)
		(56,24.642)
		(57,28.152)
		(58,22.846)
		(59,20.014)
		(60,21.268)
		(61,29.27)
		(62,25.842)
		(63,29.434)
		(64,27.73)
		(65,21.308)
		(66,30.782)
		(67,23.318)
		(68,28.248)
		(69,26.348)
		(70,27.708)
		(71,29.304)
		(72,28.406)
		(73,19.044)
		(74,24.382)
		(75,27.986)
		(76,24.198)
		(77,24.496)
		(78,28.67)
		(79,28.602)
		(80,21.602)
		(81,29.194)
		(82,30.688)
		(83,30.93)
		(84,27.984)
		(85,29.286)
		(86,27.806)
		(87,23.346)
		(88,24.19)
		(89,29.966)
		(90,22.0)
		(91,25.134)
		(92,25.48)
		(93,28.476)
	};  
	\draw [red] ({rel axis cs:0,0}|-{axis cs:93,5}) -- ({rel axis cs:1,0}|-{axis cs:93,5}) ;
	\draw [red] ({rel axis cs:0,0}|-{axis cs:93,30}) -- ({rel axis cs:1,0}|-{axis cs:93,30}) ;
	
	\nextgroupplot[bar width=0.001,
	ytick={50, 75},yticklabels={\tiny \textcolor{red}{$\tau_{low}$}, \tiny \textcolor{red}{$\tau_{high}$}},]
	\addplot coordinates {(0,53.97)
		(1,55.052)
		(2,56.216)
		(3,56.108)
		(4,66.35)
		(5,65.526)
		(6,63.558)
		(7,58.768)
		(8,61.24)
		(9,62.986)
		(10,63.124)
		(11,64.892)
		(12,64.41)
		(13,61.99)
		(14,73.1)
		(15,56.014)
		(16,63.902)
		(17,58.31)
		(18,65.248)
		(19,59.856)
		(20,66.042)
		(21,59.686)
		(22,59.528)
		(23,65.766)
		(24,66.714)
		(25,60.748)
		(26,53.012)
		(27,64.83)
		(28,59.088)
		(29,61.466)
		(30,67.778)
		(31,56.244)
		(32,64.278)
		(33,63.026)
		(34,63.566)
		(35,56.866)
		(36,63.128)
		(37,69.166)
		(38,60.81)
		(39,62.918)
		(40,65.842)
		(41,69.286)
		(42,63.284)
		(43,68.586)
		(44,60.154)
		(45,66.736)
		(46,64.068)
		(47,67.18)
		(48,61.768)
		(49,61.036)
		(50,63.384)
		(51,59.3)
		(52,66.824)
		(53,62.906)
		(54,57.972)
		(55,68.564)
		(56,56.21)
		(57,57.28)
		(58,59.954)
		(59,65.974)
		(60,60.36)
		(61,63.526)
		(62,58.318)
		(63,68.87)
		(64,57.93)
		(65,69.514)
		(66,60.676)
		(67,60.136)
		(68,60.204)
		(69,67.816)
		(70,63.05)
		(71,61.254)
		(72,66.388)
		(73,50.98)
		(74,63.544)
		(75,68.85)
		(76,71.54)
		(77,50.408)
		(78,60.316)
		(79,64.096)
		(80,64.83)
		(81,66.624)
		(82,60.76)
		(83,62.462)
		(84,59.586)
		(85,64.012)
		(86,68.114)
		(87,59.784)
		(88,63.862)
		(89,63.124)
		(90,72.424)
		(91,61.794)
		(92,64.57)
		(93,70.212)
	};  
	\draw [red] ({rel axis cs:0,0}|-{axis cs:93,50}) -- ({rel axis cs:1,0}|-{axis cs:93,50}) ;
	\draw [red] ({rel axis cs:0,0}|-{axis cs:93,75}) -- ({rel axis cs:1,0}|-{axis cs:93,75}) ;

	\nextgroupplot[bar width=0.001,
	ytick={50, 75},yticklabels={\tiny \textcolor{red}{$\tau_{low}$}, \tiny \textcolor{red}{$\tau_{high}$}},]
	\addplot coordinates {(0,67.536)
		(1,61.986)
		(2,61.922)
		(3,63.786)
		(4,62.648)
		(5,60.08)
		(6,63.136)
		(7,62.958)
		(8,63.678)
		(9,64.496)
		(10,61.21)
		(11,66.212)
		(12,73.972)
		(13,59.134)
		(14,72.108)
		(15,66.268)
		(16,63.26)
		(17,60.828)
		(18,66.58)
		(19,63.544)
		(20,64.646)
		(21,63.272)
		(22,61.678)
		(23,65.274)
		(24,71.558)
		(25,70.93)
		(26,70.744)
		(27,64.38)
		(28,59.696)
		(29,59.794)
		(30,63.81)
		(31,59.334)
		(32,59.602)
		(33,65.28)
		(34,68.06)
		(35,62.226)
		(36,70.684)
		(37,66.306)
		(38,60.518)
		(39,61.816)
		(40,61.08)
		(41,64.358)
		(42,65.388)
		(43,67.124)
		(44,66.636)
		(45,68.508)
		(46,61.706)
		(47,68.444)
		(48,66.224)
		(49,63.768)
		(50,60.958)
		(51,72.938)
		(52,64.152)
		(53,65.382)
		(54,68.108)
		(55,74.046)
		(56,60.056)
		(57,65.962)
		(58,65.768)
		(59,70.79)
		(60,64.272)
		(61,58.974)
		(62,66.434)
		(63,62.842)
		(64,65.442)
		(65,64.298)
		(66,67.58)
		(67,57.372)
		(68,63.068)
		(69,67.362)
		(70,65.926)
		(71,68.236)
		(72,71.848)
		(73,58.126)
		(74,59.818)
		(75,65.584)
		(76,61.228)
		(77,59.91)
		(78,61.47)
		(79,66.09)
		(80,63.024)
		(81,66.704)
		(82,66.49)
		(83,65.048)
		(84,58.948)
		(85,64.852)
		(86,64.858)
		(87,56.982)
		(88,59.252)
		(89,72.198)
		(90,67.004)
		(91,58.736)
		(92,64.246)
		(93,66.306)
	};  
	\draw [red] ({rel axis cs:0,0}|-{axis cs:93,50}) -- ({rel axis cs:1,0}|-{axis cs:93,50}) ;
	\draw [red] ({rel axis cs:0,0}|-{axis cs:93,75}) -- ({rel axis cs:1,0}|-{axis cs:93,75}) ;

	\end{groupplot}
	\node[text width=6cm,align=center,anchor=north] at ([yshift=-1mm]my plots c1r1.south) {{\small Class \normalsize\\ (a) UAWA-Exp1\\\normalsize\label{subplot:one}}};
	\node[text width=6cm,align=center,anchor=north] at ([yshift=-1mm]my plots c2r1.south) {{\small Class \normalsize\\ (a) NUAWA-Exp1\normalsize\label{subplot:two}}};
	\node[text width=6cm,align=center,anchor=north] at ([yshift=-1mm]my plots c3r1.south) {{\small Class \normalsize\\ (a) UAWA-Exp3\normalsize\label{subplot:three}}};
	\node[text width=6cm,align=center,anchor=north] at ([yshift=-1mm]my plots c4r1.south) {{\small Class \normalsize\\ (a) NUAWA-Exp3\normalsize\label{subplot:four}}};

	\end{tikzpicture}
	
	\caption{\footnotesize The average Bandwidth Overhead (BWO) that AWA imposes to the traces of 94 various websites for four different transformer sets.}
	\label{fig:BWOClass}
\end{figure*}
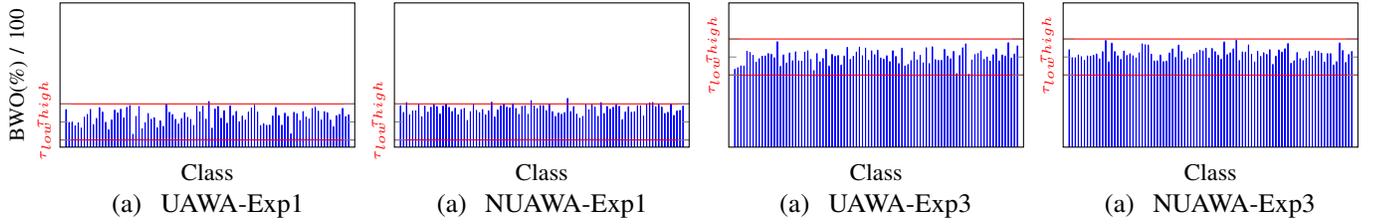

%% file: content/plots.tex
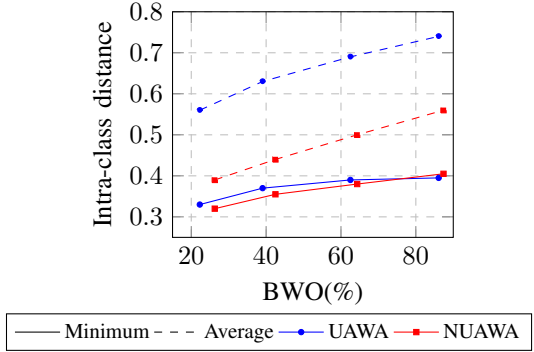
\begin{figure}[]
	\centering
	\begin{subfigure}{0.8\linewidth}
		\centering
		
		\begin{tikzpicture}
		\begin{axis}[
		xmin=15,   xmax=90,
		ymin=0.25,   ymax=0.8,
		xlabel=BWO(\%),
		ylabel=  Intra-class distance,
		grid=major,
		width=0.75\textwidth,
		legend style={at={(0.5,-0.25)},
			anchor=north,legend columns=2,font=\footnotesize},
		ytick={0.3,0.4,0.5, 0.6,0.7,0.8}
		]
		\comment{
		\addplot[color=blue,mark=*,blue,mark size=1pt,] 
coordinates { 
	(22.28 ,1.02)    
	(39.06,1.04)  
	(62.52,1.08) 
	(86.1,1.11)
	
};
\addplot[color=red,mark=square*,red,mark size=1pt] 
coordinates{ 
	(26.28, 0.79)    
	(42.49, 0.83)  
	(64.33, 0.92) 
	(87.39, 1.01)
};

		\addplot[color=blue,mark=*,blue,mark size=1pt,] 
coordinates { 
	(22.28 ,1.19)    
	(39.06,1.30)  
	(62.52,1.34) 
	(86.1,1.38)
	
};

\addplot[color=red,mark=square*,red,mark size=1pt] 
coordinates{ 
	(26.28, 1.00)    
	(42.49, 1.04)  
	(64.33, 1.09) 
	(87.39, 1.18)
};
}
\comment{
		\addplot[color=blue,mark=*,blue,mark size=1pt,] 
		coordinates { 
			(22.28 ,1.02)    
			(39.06,1.05)  
			(62.52,1.08) 
			(86.1,1.12)

		};
		}
		\addplot[color=blue,dashed,mark=*,blue,mark size=1pt,mark options={fill=blue},]
		coordinates { 
			(22.28 ,0.56)    
			(39.06,0.63)  
			(62.52,0.69) 
			(86.1,0.74)
		};
\comment{
		\addplot[color=red,mark=square*,red,mark size=1pt] 
		coordinates{ 
			(26.28, 0.80)    
			(42.49, 0.84)  
			(64.33, 0.93) 
			(87.39, 1.01)
		};
	}
		\addplot[color=red,dashed,mark=square*,red,mark size=1pt,] 
		coordinates{ 
			(26.28, 0.39)    
			(42.49, 0.44)  
			(64.33, 0.50) 
			(87.39, 0.56)
		};

		\comment{
				\addplot[color=green,mark=*,blue,mark size=1pt,] 
		coordinates { 
			(22.28 ,0.3499)    
			(39.06,0.3944)  
			(62.52,0.4283) 
			(86.1,0.46)
			
		};
	
		\addplot[color=red,dashed,mark=square*,red,mark size=1pt,] 
coordinates{ 
	(26.28,0.2409 )    
	(42.49,0.2764 )  
	(64.33,0.3104 ) 
	(87.39, 0.3471)
};
}

				\addplot[color=blue,mark=*,blue,mark size=1pt,] 
coordinates { 
	(22.28 ,0.33)    
	(39.06,0.37)  
	(62.52,0.39) 
	(86.1,0.395)
	
};

\addplot[color=red,mark=square*,red,mark size=1pt,] 
coordinates{ 
	(26.28,0.32)    
	(42.49,0.355 )  
	(64.33,0.38 ) 
	(87.39, 0.405)
};

		\end{axis}
		\end{tikzpicture}
		
		\label{fig:fcc_evala}
	\end{subfigure}
	
	\begin{subfigure}{\linewidth}
		\centering
		\begin{tikzpicture} 
		\begin{axis}[%
		hide axis,
		xmin=10,
		xmax=50,
		ymin=0,
		ymax=0.4,
		legend style={nodes={scale=0.8, transform shape},at={($(0,0)+(1cm,1cm)$)},legend columns=-1,fill=none,draw=black,anchor=center,align=center}
		]
		\addlegendimage{black}
		\addlegendentry{Minimum};  
		\addlegendimage{black,dashed}
		\addlegendentry{Average};   
		\addlegendimage{blue,mark=*,mark size=1pt,mark options={fill=blue}}
		\addlegendentry{UAWA};  
		\addlegendimage{red,mark=square*,mark size=1pt,mark options={fill=red}}
		\addlegendentry{NUAWA};

		\end{axis}
		\end{tikzpicture}
	\end{subfigure}
	\caption{ \footnotesize Minimum and average intra-class distance of UAWA and NUAWA over various sizes of Bandwidth Overhead (BWO)}
	\label{fig:intracd}
\end{figure}

%% file: content/old_ClassDistanc.tex
{\color{black}
	Since we have five transformer sets in each experiment, all websites' distribution is transformed into five different distributions. We assume an adversary and a target user each select one of these distributions, and the adversary trains a classifier on the samples of her distribution. As the distance between the adversary's distribution and the target user's distribution is increased, the accuracy of the adversary's classifier is decreased, given to the findings of \cite{DBLP:conf/iclr/ZhangCSBDH19}. We introduce Intra-Class Distance (Intra-CD) to calculate the average and minimum empirical distance between the five distributions that have been formed by five transformer sets. Intra-CD uses MMD to estimate the distance between two distributions. Since there are five different distributions, we have ten pairs of distributions.  We use Average Intra-CD (Avg Intra-CD) and Minimum Intra-CD (Min-Intra-CD) to measure the average and minimum distance between each pair of these distributions. Therefore, Avg and Min Intra-CD indicate the average and minimum distances between the distributions of adversary's adversarial traces and the target user's adversarial traces when they pick different sets of transformers.
 Avg and Min Intra-CD are defined as follows:}
\begin{equation}
\begin{split}
&MMD_k^{i,j} = MMD(T_k^i(TS_k,z),T_k^j(TS_k,z)) \\
 & \text{Avg Intra-CD} =\; \frac{1}{K}\sum_{k=1}^{K} \frac{2}{S(S-1)}\sum_{i=1}^{S-1}\sum_{j=i+1}^{S} MMD_k^{i,j}
\\
 & \text{Min Intra-CD} =\;
\frac{1}{K}\sum_{k=1}^{K} Min(\bigcup\limits_{i=1,j=1}^{S} MMD_k^{i,j} )
\end{split}
\end{equation}
where $T_k^i$ is the transformer of website $k$ in transformer set $i$, $S$ is the number of transformer sets, and $K$ is the number of classes. In our experiment, $S=5$ and $K=94$.
Figure \ref{fig:intracd} indicates Avg and Min Intra-CD of UAWA and NUAWA over various magnitudes of BWO. Avg Intra-CD of UAWA is more than NUAWA, which means that if an adversary and a target user select different sets of transformers, the average distance between the distributions of their adversarial traces is more when UAWA creates the transformers. However, Min Intra-CD of UAWA and NUAWA are very close to each other. Intra-CD demonstrates that the training data distribution of the adversary's classifier is far from the distribution of the target user's adversarial traces, and this distance is increased by adding more bandwidth overhead to traces. Therefore, the adversary's classifier is more likely to be vulnerable to the adversarial traces of the target user when their sets of transformers are different. The results also justify the better performance of AWA when bandwidth overhead is increased.

%% file: content/visualization.tex
\input{content/vis_traces}

We use the average trace as a representation of all traces of a website. The average trace is the average of a set of burst sequences over burst indexes. The first element of the average trace is the average of the first burst sizes of all traces and so on. We use the average trace to visualize the changes that AWA applies to a set of traces. Figure \ref{fig:AvgTraceViz} presents the average traces of website A ($\mu_A = Avg(TS_A)$) and website B ($\mu_B = Avg(TS_B)$) before transformation and the average traces of transformed traces of website A ($\mu_{A} = Avg(T^s_A(TS_A,z))$) and website B ($\mu^{s}_{B} = Avg(T^s_B(TS_B,z))$), where $T^s_A$ and $T^s_B$ are in the transformer set $s$. 
Each plot of Figure \ref{fig:AvgTraceViz} is from a different transformer set, and website A is website 1 in all plots. We select three transformer sets from NUMAW-Exp3, and three transformer sets UAWA-EXP3. In each transformer set, there is one website being paired by website 1 in the training phase of AWA, and we select this website as the website B in each plot of Figure \ref{fig:AvgTraceViz}.  
{\color{black}Website 1 is selected as an example, and the paired websites were determined randomly in the training phase of AWA.}
For example, the average traces of websites 1 and 38 before and after transformation are depicted in the first plot. This figure indicates that when various transformers transform the traces of website 1, the average trace of website 1 is different, which shows that the distribution of website 1 after transforming by various transformers is different. It also shows that the style of transformed traces of two websites being paired in the AWA training phase are getting close to each other.

%% file: content/vis_traces.tex
\newcommand{\fnamei}{csv/dis_pre/dis_ada1-38.csv}
\newcommand{\fnameii}{csv/dis_pre/dis_ada1-86.csv}
\newcommand{\fnameiii}{csv/dis_pre/dis_ada1-21.csv}

\newcommand{\ufnamei}{csv/udis_pre/udis_ada1-38.csv}
\newcommand{\ufnameii}{csv/udis_pre/udis_ada1-80.csv}
\newcommand{\ufnameiii}{csv/udis_pre/udis_ada1-57.csv}

\newcommand{\vcolorf}{red!30}
\newcommand{\vcolortf}{red!100}
\newcommand{\vcolors}{blue!30}
\newcommand{\vcolorts}{blue!100}
\newcommand{\vmarkf}{-}
\newcommand{\vmarktf}{*}
\newcommand{\vmarks}{|}
\newcommand{\vmarkts}{+}

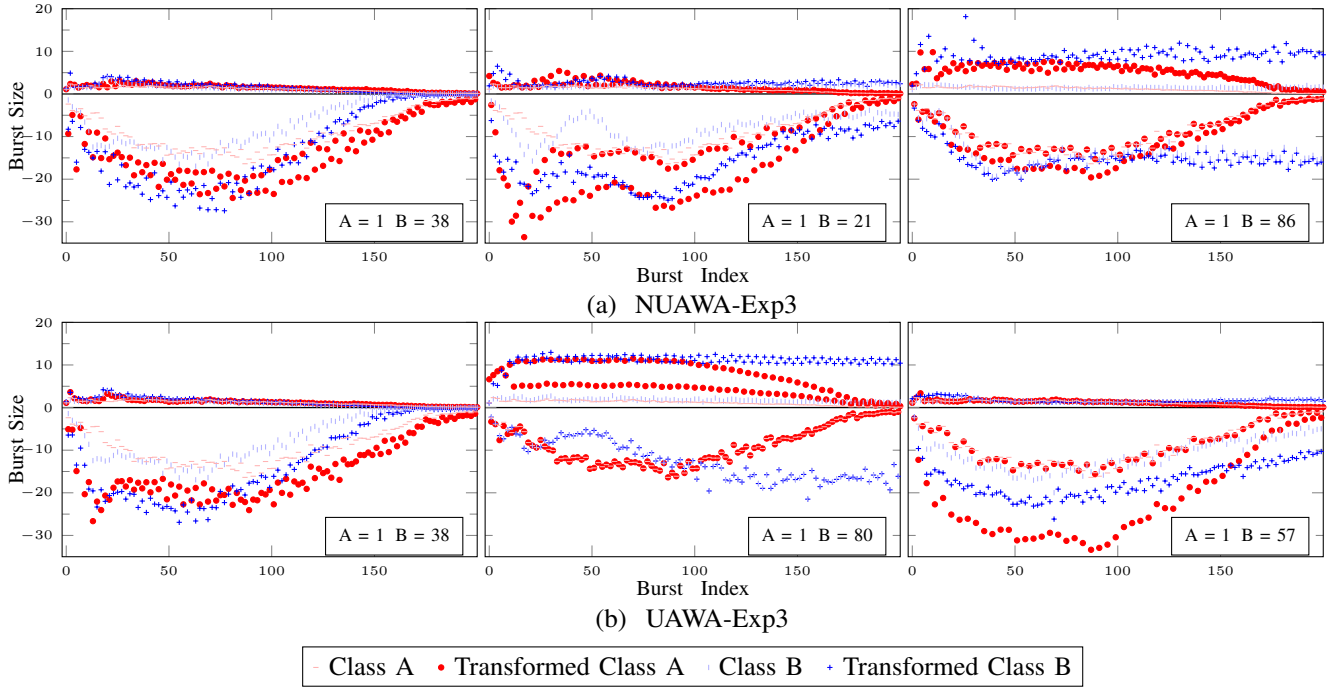
\begin{figure*}[!t]
	\centering
\begin{tikzpicture}
		\pgfplotsset{
			scaled y ticks = false,
width=7.1cm,
height=4.7cm,
axis on top,
xmin=-2,xmax=200,
ymin=-35,ymax=20,
minor y tick num=1,
label style = {font = {\fontsize{8 pt}{12 pt}\selectfont}},
title style = {font = {\fontsize{9pt}{12 pt}\selectfont}},
tick label style = {font = {\fontsize{4 pt}{12 pt}\selectfont}},
xtick={0,50,100,150},
ylabel shift={-0.8em},
ylabel style={align=center},
xlabel shift={-0.4em},
grid style={gray!10},
}
\begin{groupplot}[
group style={
	group name=my plots,
	group size=3 by 2,
	vertical sep=30pt,
	horizontal sep=3pt},
	legend style={legend columns=-1,at={(0.8,0.17)},
		anchor=north}
]
\nextgroupplot[ylabel={Burst Size},ytick={-30,-20,-10,0,10,20},]
		\addlegendimage{only marks,black, mark=.}
\addlegendentry{\scriptsize A = 1};  
		\addlegendimage{only marks,black, mark=.}
\addlegendentry{\scriptsize B = 38}; 

\addplot[only marks,mark=\vmarktf,\vcolortf,mark size=1pt,] table[x=range,y=new_src,col sep=comma]{\fnamei};\label{plots:plot2}
\addplot [solid,black, no markers,line width=0.5] coordinates {(-2,0) (200,0)};
\addplot[only marks,mark=\vmarkts,\vcolorts,mark size=1pt,] table[x=range,y=new_trg,col sep=comma]{\fnamei};\label{plots:plot4}
\addplot[only marks,mark=\vmarks,\vcolors,mark size=1pt] table[x=range,y=trg,col sep=comma]{\fnamei};\label{plots:plot3}
\addplot[only marks,mark=\vmarkf,\vcolorf,mark size=1pt] table[x=range,y=src,col sep=comma]{\fnamei};\label{plots:plot1}

\nextgroupplot[yticklabels={,,}]
		\addlegendimage{only marks,black, mark=.}
\addlegendentry{\scriptsize A = 1};  
\addlegendimage{only marks,black, mark=.}
\addlegendentry{\scriptsize B = 21}; 
\addplot[only marks,mark=\vmarktf,\vcolortf,mark size=1pt,] table[x=range,y=new_src,col sep=comma]{\fnameii};
\addplot [solid,black, no markers,line width=0.5] coordinates {(-2,0) (200,0)};
\addplot[only marks,mark=\vmarkts,\vcolorts,mark size=1pt,] table[x=range,y=new_trg,col sep=comma]{\fnameii};
\addplot[only marks,mark=\vmarks,\vcolors,mark size=1pt] table[x=range,y=trg,col sep=comma]{\fnameii};
\addplot[only marks,mark=\vmarkf,\vcolorf,mark size=1pt] table[x=range,y=src,col sep=comma]{\fnameii};
\nextgroupplot[yticklabels={,,}]
		\addlegendimage{only marks,black, mark=.}
\addlegendentry{\scriptsize A = 1};  
\addlegendimage{only marks,black, mark=.}
\addlegendentry{\scriptsize B = 86}; 
\addplot[only marks,mark=\vmarktf,\vcolortf,mark size=1pt,] table[x=range,y=new_src,col sep=comma]{\fnameiii};
\addplot [solid,black, no markers,line width=0.5] coordinates {(-2,0) (200,0)};
\addplot[only marks,mark=\vmarkts,\vcolorts,mark size=1pt,] table[x=range,y=new_trg,col sep=comma]{\fnameiii};
\addplot[only marks,mark=\vmarks,\vcolors,mark size=1pt] table[x=range,y=trg,col sep=comma]{\fnameiii};
\addplot[only marks,mark=\vmarkf,\vcolorf,mark size=1pt] table[x=range,y=src,col sep=comma]{\fnameiii};

\nextgroupplot[ylabel={Burst Size},ytick={-30,-20,-10,0,10,20},]
		\addlegendimage{only marks,black, mark=.}
\addlegendentry{\scriptsize A = 1};  
\addlegendimage{only marks,black, mark=.}
\addlegendentry{\scriptsize B = 38}; 
\addplot[only marks,mark=\vmarktf,\vcolortf,mark size=1pt,] table[x=range,y=new_src,col sep=comma]{\ufnamei};
\addplot [solid,black, no markers,line width=0.5] coordinates {(-2,0) (200,0)};
\addplot[only marks,mark=\vmarkts,\vcolorts,mark size=1pt,] table[x=range,y=new_trg,col sep=comma]{\ufnamei};
\addplot[only marks,mark=\vmarks,\vcolors,mark size=1pt] table[x=range,y=trg,col sep=comma]{\ufnamei};
\addplot[only marks,mark=\vmarkf,\vcolorf,mark size=1pt] table[x=range,y=src,col sep=comma]{\ufnamei};

\nextgroupplot[yticklabels={,,}]
		\addlegendimage{only marks,black, mark=.}
\addlegendentry{\scriptsize A = 1};  
\addlegendimage{only marks,black, mark=.}
\addlegendentry{\scriptsize B = 80}; 
\addplot[only marks,mark=\vmarktf,\vcolortf,mark size=1pt,] table[x=range,y=new_src,col sep=comma]{\ufnameii};
\addplot [solid,black, no markers,line width=0.5] coordinates {(-2,0) (200,0)};
\addplot[only marks,mark=\vmarkts,\vcolorts,mark size=1pt,] table[x=range,y=new_trg,col sep=comma]{\ufnameii};
\addplot[only marks,mark=\vmarks,\vcolors,mark size=1pt] table[x=range,y=trg,col sep=comma]{\ufnameii};
\addplot[only marks,mark=\vmarkf,\vcolorf,mark size=1pt] table[x=range,y=src,col sep=comma]{\ufnameii};

\nextgroupplot[yticklabels={,,}]
		\addlegendimage{only marks,black, mark=.}
\addlegendentry{\scriptsize A = 1};  
\addlegendimage{only marks,black, mark=.}
\addlegendentry{\scriptsize B = 57}; 
\addplot[only marks,mark=\vmarktf,\vcolortf,mark size=1pt,] table[x=range,y=new_src,col sep=comma]{\ufnameiii};
\addplot [solid,black, no markers,line width=0.5] coordinates {(-2,0) (200,0)};
\addplot[only marks,mark=\vmarkts,\vcolorts,mark size=1pt,] table[x=range,y=new_trg,col sep=comma]{\ufnameiii};
\addplot[only marks,mark=\vmarks,\vcolors,mark size=1pt] table[x=range,y=trg,col sep=comma]{\ufnameiii};
\addplot[only marks,mark=\vmarkf,\vcolorf,mark size=1pt] table[x=range,y=src,col sep=comma]{\ufnameiii};

\end{groupplot}
 \node[text width=6cm,align=center,anchor=north] at ([yshift=-2mm]my plots c2r1.south) {{\footnotesize Burst Index \normalsize\\ (a) NUAWA-Exp3\normalsize\label{subplot:two}}};
   
    \node[text width=6cm,align=center,anchor=north] at ([yshift=-2mm]my plots c2r2.south) {{\footnotesize Burst Index \normalsize\\ (b) UAWA-Exp3\normalsize\label{subplot:two}}};

   \path (my plots c2r1.north |-current bounding box.south)--
   coordinate(legendpos)
   (my plots c2r1.north |-current bounding box.south);
   \matrix[
   matrix of nodes,
   anchor=south,
   draw,
   inner sep=0.2em,
   draw
   ]at([yshift=-0.6cm]legendpos)
   {
   	\ref{plots:plot1}& Class A&[5pt]
   	\ref{plots:plot2}& Transformed Class A&[5pt]
   	\ref{plots:plot3}& Class B&[5pt]
     \ref{plots:plot4}& Transformed Class B\\};

\end{tikzpicture}

	\caption{ \footnotesize  The average traces of six pairs of websites before and after transformation. The three plots on the first and second rows are from UAWA-Exp3 and NUAWA-Exp3, respectively. In each plot, website A is website 1, and website B is the paired website with website 1 in a transformer set.}
\label{fig:AvgTraceViz}
\end{figure*}

%% file: content/SimWorks.tex
We used Deep Fingerprinting (DF) \cite{DBLP:conf/ccs/SirinamIJW18} as the adversary's classifier in the experiments. In addition to DF, there are two classifiers in the previous works which have shown higher performance in classifying user's traces in comparison to the traditional approaches. Rimmer \textit{et al.} \cite{DBLP:conf/ndss/RimmerPJGJ18} propose Automated Website Fingerprinting (AWF) classifier, and they utilize SDAE, LSTM, and CNN as the adversary's classifier. AWF uses the sequence of packets direction to classify traces. Bhat \textit{et al.} \cite{DBLP:journals/popets/BhatLKD19} present VAR-CNN and use the sequence of packets direction, inter-arrival time, and cumulative statistical features to classify traces.  The baseline CNN architecture of VAR-CNN is based on ResNet-18 \cite{7780459}. VAR-CNN runs in 150 epochs and uses learning rate decay and early-stopping to improve the classifier's performance.
Table \ref{tab:otherattack} indicates the performance of UAWA-Exp1 and NUAWA-Exp1 against AWF and ResNet-18 (VAR-CNN) in Scenario 1. We use the CNN classifier of AWF, and since our dataset only has the sequence of packets direction, we train a ResNet-18 classifier on them for VAR-CNN. The results demonstrate that AWA is effective against all three attacks.
\begin{table}[]
	\centering
	\caption{ \footnotesize The average and standard deviation of the accuracy of various attacks that have been proposed in previous studies \cite{DBLP:conf/ccs/SirinamIJW18,DBLP:conf/ndss/RimmerPJGJ18,DBLP:journals/popets/BhatLKD19} against UAWA-Exp1 and NUAWA-Exp1.}
	\label{tab:otherattack}
			\resizebox{0.50\textwidth}{!}{ 
\begin{tabular}{clcclcc}
	&  & \multicolumn{2}{c}{UAWA-Exp1}   &  & \multicolumn{2}{c}{NUAWA-Exp1}  \\ \cline{3-4} \cline{6-7} 
	&  & \multicolumn{2}{c}{22.28\% BWO} &  & \multicolumn{2}{c}{26.28\% BWO} \\ \cline{3-4} \cline{6-7} 
	Adversary's name    &  & Avg Acc(\%)    & STD Acc(\%)    &  & Avg Acc(\%)     & STD Acc(\%)   \\ \hline
	DF \cite{DBLP:conf/ccs/SirinamIJW18}                  &  & 19.52         & 2.34          &  & 31.94           & 2.71        \\
	AWF \cite{DBLP:conf/ndss/RimmerPJGJ18}         &  &    15.19       & 2.29         &  & 18.87           & 1.32        \\
	ResNet-18 (VAR-CNN) \cite{DBLP:journals/popets/BhatLKD19} &  & 16.80          & 1.96         &  & 26.60           & 2.40    
\end{tabular}
}
\end{table}

\begin{table}[]
	\centering
	\caption{ \footnotesize The accuracy of Deep Fingerprinting (DF) attack \cite{DBLP:conf/ccs/SirinamIJW18} against AWA, Mockingbird, and WTF-PAD.}
	\label{tab:otherdefense}
	\resizebox{0.35\textwidth}{!}{ 
		\begin{tabular}{ccclc}
			Defense name        &        & BWO(\%)       &        & DF accuracy(\%)       \\ \hline
			Mockingbird  \cite{imani2019mockingbird}       &        & $58.0^*$          &        & $42.0^*$              \\
			WTF-PAD  \cite{DBLP:conf/esorics/JuarezIPDW16}           &        & $64.0^*$          &        & $86.0^*$                  \\ \cline{1-1} \cline{3-3} \cline{5-5} 
			UAWA-Exp1           &        & 22.28         &        & 19.52                 \\
			NUAWA-Exp1          &        & 26.28         &        & 31.94                 \\ \hline
			\multicolumn{5}{l}{$*$ are from \cite{imani2019mockingbird}}
		\end{tabular}
	}
\end{table}

We compare the performance of AWA with Mockingbird \cite{imani2019mockingbird} and WTF-PAD \cite{DBLP:conf/esorics/JuarezIPDW16} in Table \ref{tab:otherdefense}.
Since Mockingbird requires to have access to the entire trace of a website before generating an adversarial example, it is not universal. WTF-PAD \cite{DBLP:conf/esorics/JuarezIPDW16} is the most promising defense that does not use adversarial machine learning techniques. Saidur \textit{et al.} in \cite{imani2019mockingbird} have reported the performance of Mockingbird and WTF-PAD against DF attack on the same dataset that we use in this study.
Table \ref{tab:otherdefense} indicates the performance of Mockingbird, WTF-PAD, UAWA-Exp1, and NUAWA-Exp1 against DF attack. The results demonstrate that AWA is more effective than Mockingbird and WTF-PAD defenses with lower BWO.

%% file: content/AWAPrac.tex
AWA has three phases: pre-training, training, and testing. The pre-training and training phases can be done offline. The generation process of adversarial traces (testing phase) in UAWA also can be done offline, but in NUAWA, it should be done online. The training process of Deep Neural Networks (DNNs) is often done on Graphical Process Units (GPUs) due to their performance. GPUs reduce the training and testing time by orders of magnitude due to their much more efficient matrix operations.
 Hence, the pre-training and training phases of AWA should be done on GPUs. It is unrealistic to suppose that all users have access to GPUs. Therefore, PETs should provide a GPU server for three phases of AWA. This server is called  AWA server. Nevertheless, if a user has access to GPU, she can run AWA on her GPU. The AWA server must run AWA multiple times with various random elements to create multiple sets of transformers. Since AWA does not need random elements after the training phase, except for the noise input of generators in UAWA, all other parts of secret random elements can be removed after the AWA training phase. The noise input of generators in UAWA also can be removed after generating perturbation vectors. The AWA server must save created transformers sets on a secure storage. 
	Notably, the computational cost of the AWA training phase linearly increases by increasing the number of websites. However, it does not affect the computational cost of the testing phase.

When a user for the first time uses AWA, a transformer set is assigned to that user. There are two solutions for a user to generate adversarial traces. First, she can download transformers and use them on her local machine. The size of each transformer, which is equal to the size of generator parameters, is about 112 KB. The user can utilize a Core Process Unit (CPU) or GPU to generate adversarial traces using downloaded transformers. Generating 100 adversarial traces takes about 0.02 seconds on GPU (GeForce GTX 1080 Ti) and about 0.35 seconds on CPU (Intel(R) Core(TM) i7-4710HQ). In the second solution, which is more convenient, the AWA server is responsible for generating adversarial traces. In UAWA, for each user, the AWA server can generate perturbation vectors (the output of the generator) of various websites through the associated transformer set and send them to the user. In this setting, when a user wants to visit website A, she can use the perturbation vector of website A downloaded beforehand. The elements of perturbation vector determine how many dummy packets must be added to the end of each burst on the fly. Therefore, UAWA has no computation cost in the browsing time. In NUAWA, the entire trace of a website must be fed to the transformer. Therefore, similar to \cite{imani2019mockingbird,DBLP:conf/uss/WangCNJG14}, there must be a database of fresh traces of various websites in the AWA server. In this setting, when a user wants to visit website A, the AWA server using the associated transformer generates an adversarial trace for website A and send it to her. The user must send real packets and dummy packets based on the received adversarial trace. Hence, there is a little computation cost for the AWA server to generate adversarial traces in NUAWA.

AWA injects dummy packets to both sides of network traffic, Client to Server (C2S) and Server to Client (S2C). A user can send C2S dummy packets by her machine. The entity that sends S2C dummy packets depends on the infrastructure of PETs. A web server or a middleware can send S2C dummy packets. In Tor, S2C dummy packets can be sent by the first node (bridge node) of the Tor network. Hence, adversarial trace in NUAWA and perturbation vector in UAWA should also be sent to the bridge node.

%% file: content/conclusion.tex
In this paper, we proposed a novel defense against the website fingerprinting attack called Adversarial Website Adaptation (AWA), which has two versions, Universal AWA (UAWA) and Non-Universal AWA (NUAWA). We considered two scenarios to evaluate the performance of AWA. In the first scenario, we have shown that if an adversary and a target user generate their traces by different sets of transformers, both UAWA and NUAW are highly effective; however, the performance of UAWA is better. In the second scenario, we have indicated that if an adversary uses multiple sets of transformers that are different from the target user's set of transformers, NUAWA is much more effective than UAWA. The results demonstrate when the adversary is more powerful and can collect traces of various websites through multiple sets of transformers, the defense can not use the benefits of UAWA and must impose more bandwidth overhead to traces. 

The future work is to provide theoretical analysis for AWA using several recent works \cite{8835364} on certified robustness against adversarial attacks that can provide formal guarantees of upper and lower bound for required noise.

%% file: content/Apendix_A.tex
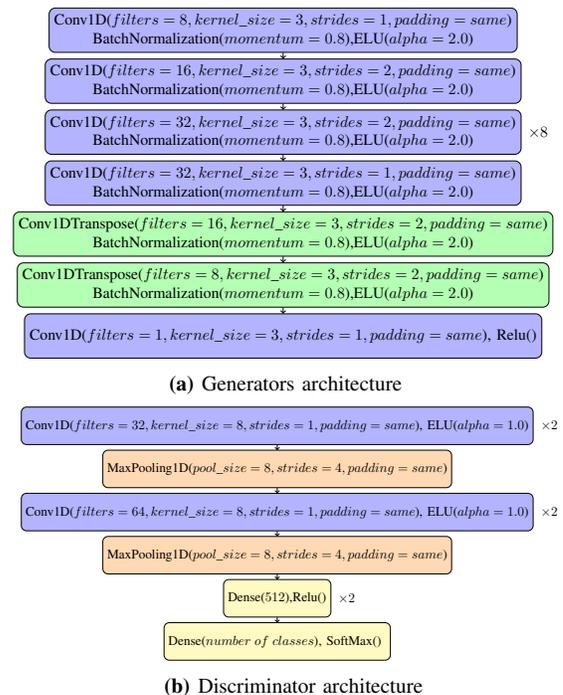
\begin{figure}[H]
	\centering
	\begin{subfigure}[H]{0.4\textwidth}
		\centering
		\resizebox{1\textwidth}{!}{%
\begin{tikzpicture}[node distance=1.15cm,
every node/.style={fill=white}, align=center]
\node (start)             [conv]              {Conv1D($filters=8, kernel\_size=3, strides=1, padding=same$)\\BatchNormalization($momentum=0.8$),ELU($alpha=2.0$)};
\node (onCreateBlock)     [conv, below of=start]          {Conv1D($filters=16, kernel\_size=3, strides=2, padding=same$)\\BatchNormalization($momentum=0.8$),ELU($alpha=2.0$)};
\node (onStartBlock)      [conv, below of=onCreateBlock]   {Conv1D($filters=32, kernel\_size=3, strides=2, padding=same$)\\BatchNormalization($momentum=0.8$),ELU($alpha=2.0$)};
\node (onResumeBlock)     [conv, below of=onStartBlock]   {Conv1D($filters=32, kernel\_size=3, strides=1, padding=same$)\\BatchNormalization($momentum=0.8$),ELU($alpha=2.0$)};
\node (txtnod) [right of = onStartBlock,xshift=4.6cm] { $\times 8$};
\node (activityRuns)      [deconv, below of=onResumeBlock] {Conv1DTranspose($filters=16, kernel\_size=3, strides=2, padding=same$)\\BatchNormalization($momentum=0.8$),ELU($alpha=2.0$)};
\node (onPauseBlock)      [deconv, below of=activityRuns]
{Conv1DTranspose($filters=8, kernel\_size=3, strides=2, padding=same$)\\BatchNormalization($momentum=0.8$),ELU($alpha=2.0$)};
\node (onStopBlock)       [conv, below of=onPauseBlock]
{Conv1D($filters=1, kernel\_size=3, strides=1, padding=same$), Relu()};

\draw[->]             (start) -- (onCreateBlock);
\draw[->]     (onCreateBlock) -- (onStartBlock);
\draw[->]      (onStartBlock) -- (onResumeBlock);
\draw[->]     (onResumeBlock) -- (activityRuns);
\draw[->]      (activityRuns) --  (onPauseBlock);
\draw[->]      (onPauseBlock) --(onStopBlock);

\end{tikzpicture}
		}%
		\caption{Generators architecture\\$\;$}
	\end{subfigure}%
\vspace{-0.2cm}
	\begin{subfigure}[t]{0.5\textwidth}
	\centering
	\resizebox{0.8\textwidth}{!}{%
\begin{tikzpicture}[node distance=1.15cm,
every node/.style={fill=white}, align=center]
\node (start)             [conv]              {Conv1D($filters=32, kernel\_size=8, strides=1, padding=same$), ELU($alpha=1.0$)};
\node (txtnod) [right of = start,xshift=6.1cm] { $\times 2$};
\node (onStartBlock)      [maxpool, below of=start]   {MaxPooling1D($pool\_size=8, strides=4,padding=same$)};
\node (onResumeBlock)     [conv, below of=onStartBlock]   {Conv1D($filters=64, kernel\_size=8, strides=1, padding=same$), ELU($alpha=1.0$)};
\node (txtnod2) [right of = onResumeBlock,xshift=6.1cm] { $\times 2$};
\node (onPauseBlock)      [maxpool, below of=onResumeBlock]
{MaxPooling1D($pool\_size=8, strides=4,padding=same$)};
\node (onStopBlock)       [dense, below of=onPauseBlock]
{Dense($512$),Relu()};
\node (txtnod32) [right of = onStopBlock,xshift=.7cm] { $\times 2$};
\node (onStopBlock2)       [dense, below of=onStopBlock]
{ Dense($number\; of\; classes$), SoftMax() };

\draw[->]             (start) -- (onStartBlock);
\draw[->]      (onStartBlock) -- (onResumeBlock);
\draw[->]     (onResumeBlock) -- (onPauseBlock);
\draw[->]      (onPauseBlock) --(onStopBlock);
\draw[->]      (onStopBlock) --(onStopBlock2);

\end{tikzpicture}
		}%
		\caption{Discriminator architecture}
	\end{subfigure}
	\caption{The DNN architectures of generator and discriminator in the AWA framework.}
	\label{FIG:AWA}
\end{figure}